\documentclass[11pt,a4paper]{article}

\usepackage[margin=1in]{geometry}
\usepackage{amsmath,amssymb,amsfonts}
\usepackage{graphicx}
\usepackage{authblk}
\usepackage{hyperref}
\usepackage{caption}
\usepackage{float}
\usepackage[numbers]{natbib}

\usepackage{graphicx} 
\usepackage{algorithm}
\usepackage{algpseudocode}

\usepackage{caption,tabularx,booktabs}
\usepackage{subcaption}

\usepackage{booktabs}
\usepackage{ upgreek }

\usepackage{float} 

\usepackage{listings}
\usepackage{color} 

\usepackage{listings}
\definecolor{codegreen}{rgb}{0,0.6,0}
\definecolor{codegray}{rgb}{0.5,0.5,0.5}
\definecolor{codepurple}{rgb}{0.58,0,0.82}
\definecolor{backcolour}{rgb}{0.95,0.95,0.92}

\lstdefinestyle{mystyle}{
    backgroundcolor=\color{backcolour},   
    commentstyle=\color{codegreen},
    keywordstyle=\color{magenta},
    numberstyle=\tiny\color{codegray},
    stringstyle=\color{codepurple},
    basicstyle=\ttfamily\footnotesize,
    breakatwhitespace=false,         
    breaklines=true,                 
    captionpos=b,                    
    keepspaces=true,                 
    numbers=left,                    
    numbersep=5pt,                  
    showspaces=false,                
    showstringspaces=false,
    showtabs=false,                  
    tabsize=2
}

\title{Exploring the Role of Theory of Mind in Human Decision Making: Cognitive, Spatial, and Emotional Influences in the Adversarial Rock-Paper-Scissors Game}
\author[1]{Thuy Ngoc Nguyen}
\author[2]{Jeffrey Flagg}
\author[2]{Cleotilde Gonzalez}
\affil[1]{University of Dayton, 300 College Park, Dayton, 45469, Ohio, USA}
\affil[2]{Carnegie Mellon University, 5000 Forbes Ave, Pittsburgh, 15213, Pennsylvania, USA}
\date{Preprint}

\begin{document}
\maketitle

\begin{abstract}
Understanding how humans attribute beliefs, goals, and intentions to others—known as theory of mind (ToM)— is critical in the context of human-computer interaction. Despite various metrics used to assess ToM, the interplay between cognitive, spatial, and emotional factors in influencing human decision making during adversarial interactions remains underexplored. This paper investigates these relationships using the Rock-Paper-Scissors (RPS) game as a testbed. Through established ToM tests, we analyze how cognitive reasoning, spatial awareness, and emotional perceptiveness affect human performance when interacting with bots and human opponents in repeated RPS settings. Our findings reveal significant correlations among certain ToM metrics and highlight humans' ability to detect patterns in opponents' actions. However, most individual ToM metrics proved insufficient for predicting performance variations, with recursive thinking being the only metric moderately associated with decision effectiveness. Through exploratory factor analysis (EFA) and structural equation modeling (SEM), we identified two latent factors influencing decision effectiveness: Factor 1, characterized by recursive thinking, emotional perceptiveness, and spatial reasoning, positively affects decision-making against dynamic bots and human players, while Factor 2, linked to interpersonal skills and rational ability, has a negative impact. These insights lay the groundwork for further research on ToM metrics and for designing more intuitive, adaptive systems that better anticipate and adapt to human behavior, ultimately enhancing human-machine collaboration.

\end{abstract}



\vspace{1em}

\maketitle

\section{Introduction}
\label{sec:introduction}

Theory of Mind (ToM) refers to the ability to understand and infer the mental states of others, including their knowledge, emotions, and intentions, which are crucial for human social interactions. This capability develops at an early stage in life and is applicable in various situations. ToM has philosophical and psychological foundations and has been the subject of extensive research \cite{scassellati2002theory, schaafsma2015deconstructing}. Despite extensive research, there remains a lack of consensus on the precise definition and measurement of ToM \cite{apperly2012theory, eddy2019you, beaudoin2020systematic, quesque2020theory, schaafsma2015deconstructing}. This gap in understanding is crucial, as accurate measurement of ToM is essential to advance research in both human and artificial intelligence (AI) contexts, where improving interactions between AI systems and humans is increasingly important \cite{li2021human,strachan2024testing}.


Research on ToM ability in humans has progressed markedly, including neuroimaging studies that have identified specific brain regions involved in ToM activation \cite{schurz2014fractionating} and investigations across various vulnerable groups such as children, individuals with autism, dementia, brain injuries, and schizophrenia, all of which demonstrated ToM impairments \cite{wellman2001meta, senju2009mindblind, cuerva2001theory, apperly2004frontal, fretland2015theory}. These studies highlight the importance of ToM in understanding cognitive and social functions and emphasize the need for clear and precise methods of measuring it.

The operationalization of ToM has been long debated, particularly regarding the distinction between explicit and implicit processing of mental states \cite{frith2008implicit, apperly2009humans}. Cognitive components play a prominent role in explicit ToM processing, as studies have shown that performance in ToM tasks is impaired when cognitive resources are taxed \cite{apperly2006belief}. Moreover, many widely used ToM tests for humans focus on the attribution of beliefs through recursive thinking \cite{o2015ease}, understanding false beliefs \cite{baron1999recognition}, and strange stories \cite{happe1994advanced}; all require significant cognitive processing \cite{webb2023emergent,strachan2024testing,han2024inductive}. In contrast, facets related to empathy and emotion recognition are often considered implicit processing and distinct from ToM. Neuroimaging studies show that these emotional processes activate different brain regions than those used for cognitive ToM tasks \cite{schurz2021toward, schurz2015evaluation}. Conversely, other studies suggest that social perception, action observation, and cognitive processing are connected through the posterior superior sulcus, promoting functional connectivity between these components \cite{yang2015integrative}. Although fMRI studies highlight regional differences between cognitive, emotional, and spatial tasks \cite{schaafsma2015deconstructing}, they reveal significant overlap in brain activation during ToM tasks \cite{denny2012meta, carrington2009there}. A meta-analysis confirms this overlap and specialization in brain functioning during ToM tasks \cite{schurz2014fractionating}. 

Despite the mixed results surrounding the operationalization of ToM, recent studies on ToM abilities in humans indicate that while cognitive evaluations are important, they primarily focus on particular aspects of ToM and may not fully assess all relevant mental states, including spatial reasoning and emotional perceptiveness \cite{shapira2023clever, sap2019socialiqa, ma2023towards}. Previous research has highlighted the role of affective perspective-taking, which involves understanding emotions and interpersonal perceptions that influence individual beliefs and behaviors \cite{borke1971interpersonal,engel2014reading,pons2019test}. Furthermore, evidence shows that spatial reasoning is integral to ToM, especially in traditional abstract reasoning tests \cite{webb2023emergent}. This type of reasoning, particularly motion perception, is associated with ToM development in children \cite{rice2016biological, miller2013individual}. In adults, engaging with others' visuospatial perspectives activates specific areas of the brain during ToM processing, unlike when focusing solely on one’s self-perspective \cite{gunia2021brain}. In addition, understanding spatial relationships is essential for planning actions based on the intentions of another agent \cite{ho2022planning}. 
These insights highlight the interaction among the cognitive, emotional, and spatial components for understanding and interacting with others.

In terms of applications, numerous studies have demonstrated the role of ToM capabilities in predicting others' actions, strategic planning, and various aspects of social reasoning and decision making \cite{ho2022planning, pereira2016integrating, rusch2020theory, baker2017rational, lim2020improving}. These abilities are crucial in predicting competitive behaviors and strategies within adversarial settings, which have wide-ranging applications in security and defense \cite{fisher2008rock}. However, applying ToM reasoning in such contexts, where one person seeks to gain at another's expense, is particularly challenging \cite{de2018estimating, zhang2021rock, brockbank2024repeated}. Specifically, the complexity of infinitely recursive guessing about an opponent's beliefs necessitates an exploration of ToM's role in adversarial behavior \cite{frey2013cyclic}. Importantly, previous studies have emphasized the adaptability of human decision behavior in response to different opponent strategies \cite{moisan2017security}, but it is not yet clear how these adaptive abilities are associated with ToM.

In general, the literature indicates that cognitive, social, emotional, and spatial processes are crucial for a robust ability to understand the beliefs, desires, and actions of others. However, little is currently understood about the relationship between these elements, whether they are part of a single construct or if individual elements can predict human behavior in situations requiring ToM. To this end, our work investigates how these cognitive, spatial, and emotional perceptiveness metrics impact human decision making in human-AI interactions where people were paired with different bots and human opponents during repeated Rock-Paper-Scissors (RPS) games with imperfect information. The repeated RPS game, a mixed strategy equilibrium scenario, provides an ideal paradigm for studying how people adapt to an opponent's sequential actions \cite{brockbank2021formalizing, guennouni2022transfer}. 

Our research sheds light on the interplay among the cognitive, emotional, and spatial components of ToM. We also explore their interconnected aspects in predicting human decision behavior in adversarial settings, which enhances our understanding of ToM assessment. 
Specifically, our research aims to address the following research questions:

\begin{itemize}
    \item \textbf{RQ1}: How are the Theory of Mind (ToM) metrics in terms of cognitive, spatial, and emotional perceptiveness correlated with each other?
    \item \textbf{RQ2}: How do human players adjust their strategies when competing against bots versus other humans in a repetitive adversarial game?
    \item \textbf{RQ3}: How does measuring human ToM ability, assessed through ToM metrics, impact their performance in repeated RPS games?
    \item \textbf{RQ4}: To what extent different ToM metrics are different terms for the same construct, and to what extent can they affect human decision performance in repeated RPS play?
    
\end{itemize}

\section{Related Work}
\label{sec:related_work}
\subsection{Measurement of Theory of Mind in Humans}


Research has explored various metrics to assess ToM abilities in humans, focusing on the attribution of beliefs, intentions, abstract reasoning with spatial awareness, and emotional understanding \cite{yoshida2008game, grant2017can, sap2019socialiqa}. These investigations mainly use a variety of cognitive and reasoning tasks to assess the complexity of human mental processes and to understand the role of ToM in cognitive and social functions \cite{strachan2024testing}.

In particular, \cite{trott2023large} and \cite{kosinski2023theory} examined the attributions of false beliefs~\cite{wimmer1983beliefs}. Complementing these studies, \cite{webb2023emergent} applied text-based matrix reasoning tasks inspired by Raven's Standard Progressive Matrices, traditionally used to measure fluid intelligence, to evaluate abstract reasoning abilities, including spatial awareness. \cite{han2024inductive} further investigated inductive reasoning through induction tasks, comparing logical reasoning processes. Similarly, \cite{strachan2024testing} conducted a series of ToM tests focusing on cognitive reasoning, including the hinting task \cite{corcoran2003inductive}, the false belief task \cite{wimmer1983beliefs}, the recognition of faux pas \cite{baron1999recognition}, and strange stories \cite{happe1994advanced} to assess humans' ability to recognize belief, intention, and desire. However, studies by \cite{ullman2023large} and \cite{shapira2023clever} suggest that while these cognitive reasoning tests are specific, they may not assess all relevant mental states in a comprehensive way, such as emotions and social perceptiveness.

Essentially, ToM encompasses a wide range of concepts, including cognitive and affective attributions, trait judgments, emotional recognition, empathy, emotional reactivity, behavioral prediction, spatial recognition, and perspective taking~\cite{milcent2022using,mallick2024you,rapp2018designing}. A meta-analysis of fMRI data supports a significant overlap in brain activation during ToM tasks, along with specialization between cognitive, emotional and spatial tasks \cite{schurz2014fractionating}. ToM research has also highlighted that ToM ability extends beyond cognitive reasoning to include emotional intelligence and social perceptiveness, often measured using the Reading the Mind in the Eyes (RME) test \cite{baron2001reading, engel2014reading}. Studies indicate that average RME scores positively correlate with collective intelligence, making it a strong predictor of group performance in various tasks \cite{woolley2010evidence, almaatouq2024effects}. Moreover, working memory for facial emojis is closely related to individual ToM ability measured by the RME social perceptiveness test, suggesting a link to social aspects of facial expressions \cite{li2024processing,lee2020perceiving}. Other ToM studies have also explored interpersonal perception \cite{borke1971interpersonal} and morally related emotions \cite{pons2019test} to assess the perception of affective perspectives. For example, the interaction between low ToM abilities and high levels of empathy can affect the effectiveness of prosocial applications in promoting prosocial behaviors \cite{shoshani2022tablet,shoshani2024impact}.

Based on the literature, we identify three conceptual factors that influence ToM: cognitive, emotional, and spatial reasoning. Cognitive reasoning involves rational and analytical processing, such as understanding false beliefs, recursive thinking, and attributing intentions. Emotional reasoning captures affective processing, including social perceptiveness and empathy. Spatial reasoning involves interpreting purposeful movements to understand direction, location, and actions, as exemplified by perspective taking and orientation tasks. We also observe that existing ToM evaluations have yet to explore their interrelationships or whether they form a single construct. This work aims to fill this gap by investigating the interconnections of these different dimensions of ToM, assessed through various metrics, to determine whether they represent different terms of the same construct.

\subsection{Predicting Human Behavior in repeated Rock Paper Scissor games} 


ToM studies have increasingly focused on dynamic decision making in adversarial settings, particularly through the well-known Rock Paper Scissors (RPS) game \cite{fisher2008rock}. RPS is a two-player, symmetric zero-sum game in which each player simultaneously selects one of three actions: ``rock'', ``paper'' or ``scissors''. If both players choose the same action, the game ends in a draw. Otherwise, ``rock'' beats ``scissors'', ``scissors'' beats ``paper'', and ``paper'' beats ``rock''. Known for its unique mixed-strategy equilibrium (MSE), RPS requires players to choose each option with equal probability to avoid predictability and exploitation by opponents \cite{de2012higher, brockbank2024repeated}. This game is particularly prominent for investigating players' ability to predict and counter their opponent's choices, which involve anticipating others' actions, a core aspect of ToM~\cite{batzilis2019behavior,zhang2021rock}.

Much research has focused on investigating the dynamics of players' strategies in repetitive RPS settings, wherein data were collected from multiple trials of repeated pair interactions \cite{hoffman2015experimental, zhang2020rock}.
For instance, \cite{batzilis2019behavior,de2013much} explored the use of cognitive resources and recursive thinking by players to strategically predict and counteract opponents' moves. Their computational simulations demonstrated that players with higher-order ToM capabilities, including the ability to infer others' intentions, outperform those with lower-order ToM in RPS games. In another work, players are observed to often adopt a "win-stay, lose-shift" strategy, where successful moves are repeated and unsuccessful ones are changed \cite{dyson2016negative, wang2014social}. \cite{forder2016behavioural} further implied that the actions ``win-stay'' are associated with the more rational and strategic System 2 processes, whereas the actions ``lose-shift'' align with the intuitive reactions of System 1 \cite{kahneman2011thinking}. 


To investigate adaptive reasoning about sequential opponent behavior, prior research usually employs algorithmic bot opponents with specific strategies, such as static or dynamic patterns, to test individuals' ability to recognize and exploit these patterns \cite{dyson2018failure, moisan2017security}. \cite{zhang2021rock} and \cite{brockbank2024repeated} examined human behavior against human and algorithmic bot opponents, with bots varying in complexity from static ``win-stay, lose-shift'' strategies to more dynamic patterns with probabilistically determined moves. Their findings indicate that individuals can exploit simple and recognizable patterns, but often struggle against more complex strategies. This highlights the need to examine how ToM abilities affect human players' strategic adaptations and their ability to identify and exploit a broad range of opponent patterns, which remain unanswered by previous studies.

To address this gap, this work studies the relationship between the ToM abilities of the players, characterized by cognitive, spatial, and emotional aspects, and their decision behavior. Specifically, we investigate the impact of ToM abilities on players' adaptive behavior, measured by decision effectiveness, and their ability to recognize and exploit opponents, measured by prediction accuracy. This investigation focuses on the RPS game against opponents with increasing levels of strategic complexity: bots with static preferred transitions, bots with dynamic patterns, and human players, who are anticipated to employ the most dynamic strategies.

\section{Methods}
\label{sec:study_design}
Our research aims to examine the relationship between ToM ability, characterized by cognitive, emotional, and spatial reasoning metrics, and their effects as direct predictive factors on human decision-making in a sequential adversarial RPS game, where predicting the opponent's move (i.e., ToM ability) is crucial. In the following section, we outline our experimental design for three studies in which participants engage in tasks involving different types of opponents, such as bots and humans, situating our work within the context of human-AI decision-making tasks.

More concretely, we investigate human performance in repeated RPS interactions against different bots and human opponents to examine how these interactions are associated with their ToM ability, measured by cognitive, spatial reasoning, and social-emotional perceptiveness. These studies are motivated by previous research to understand the impact of dynamic strategies of opponents on human adaptive reasoning \cite{zhang2021rock, brockbank2024repeated}. In Experiment 1, participants were matched with bot opponents who followed specific stable patterns in their moves. In Experiment 2, participants competed against bot opponents using stochastic conservative strategies, while in Experiment 3, they were paired with human opponents.

\subsection{General Study Design}
\begin{figure}[!htpb]
\vspace{-1ex}
\begin{center}
\includegraphics[width=.89\textwidth]{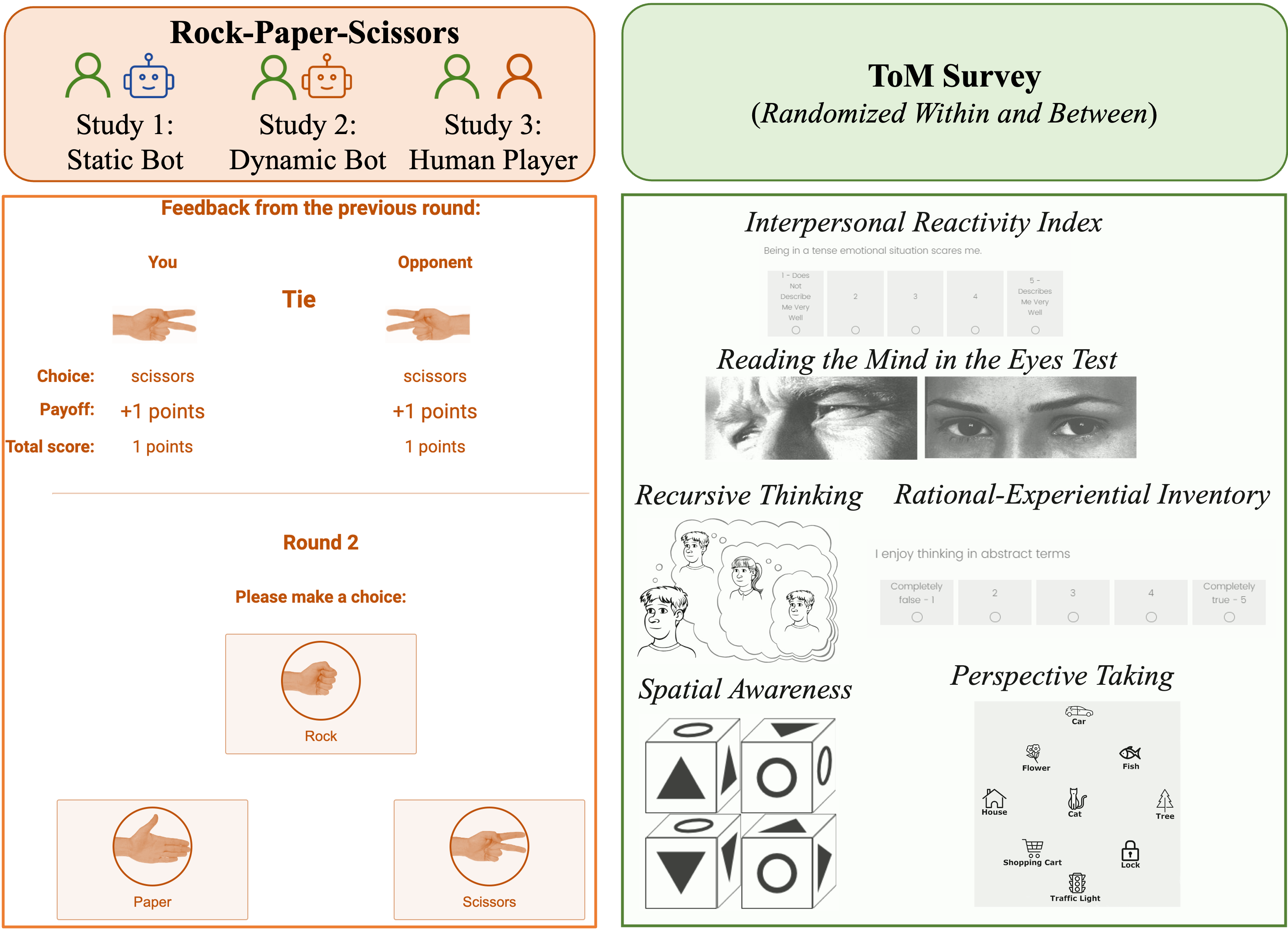} 
\caption{Schematic illustration of the study design.}
\label{fig:study-design}

\end{center}
\vspace{-1ex}
\end{figure}

Fig.\ref{fig:study-design} illustrates the general design template, followed by the three studies. The deviations from this template and the exact details unique to each study are noted in the procedure section of the particular study below. 

All participants were recruited through Amazon Mechanical Turk (MTurk). Participants received four dollars for completion of the study. 
We only included participants who were located in the United States and had an Amazon MTurk HIT approval rating of at least 90\%.

Participants begin by completing a consent form and some brief demographic questions before receiving detailed study instructions. The participants then completed two blocks, one is the RPS block, and the other is the ToM Survey. 

Within the ToM Survey, the order of all questionnaires was randomized. Furthermore, the order of the questions in each questionnaire was randomized, as was the order of the answers. Finally, two attention checks were included during the surveys. One attention check appeared during the Reading the Mind in the Eyes portion, and the other appeared in the Perspective Taking portion. In each, participants were told that this was an attention check and were instructed to select one particular answer from the selected choices. If participants failed both, they were removed from the study. After completing both blocks, the participants received a completion code and finished the study.

Fig.\ref{fig:study-design} illustrates screenshots of the experiment; and see Sections~\ref{subsubsect:tom-survey} and \ref{subsubsect:rps} for more details on each experiment.

\subsection{A Comprehensive ToM Survey}
\label{subsubsect:tom-survey}

We adopted several questionnaires that capture different aspects of ToM, classified as cognitive, emotional, and spatial factors. Within these groups, we searched for questionnaires that were widely used and validated. Studies were excluded if they were not appropriate for typical adult populations. Not every questionnaire was specifically designed to test ToM; however, we attempted to select studies that would assess important components that contribute to ToM processing. The questionnaires were considered if they were practical for implementation in online environments. They should be easily created using basic survey tools such as Qualtrics or SurveyMonkey and do not involve any special skills or time-consuming programming. Participants should be able to respond without installing any software or making audio or video recordings. Participant responses should also be easily scored and coded and do not involve open questions that must be manually coded by the experimenter. Ultimately, the goal would be to create a survey that would allow participants' scores to be used in real time during a study. The order of these questions was randomized when presented to the participants. The complete set of these ToM questionnaires is provided in the Appendix~\ref{appendx:tom-survey}.

\subsubsection{Cognitive Reasoning}
The \textit{Rational-Experiental Inventory} (REI) measures two modes of thinking, the rational information processing system and the experiential processing system \cite{pacini1999relation}. Those showing more rational processing have a tendency for deliberative, conscious, emotion-fee thinking. Those who show a preference for the experiential processing system favor automatic, unconscious, and emotional thinking. 

In this survey, there are 40 questions in total, 20 for each subsection. Participants are asked to read questions about their thinking styles. Each question ranges from 1-5, from Completely False to Completely True. The maximum on each subscale would be 100 with a minimum of 20. We expect that those who show a tendency for rational processing should be able to better think about others' mindsets and take their perspective.

The \textit{Recursive Mindreading Questionnaire} involves stories developed by \cite{o2015ease}. Each story describes a complex social situation and is followed by five recursive reasoning questions. The questions ranged from one first-order question (e.g., Person A believes Z) to a fifth-order recursive thinking question (e.g., Person X believes that Person B thinks that Person C believes that Person D knows that Person E understands Z). The maximum score was 20 and the minimum was 0. Other recursive thinking tasks are available; however, many do not contain as many levels of recursive thinking and are more difficult to implement in the form of a questionnaire.

\subsubsection{Emotional Reasoning}

The \textit{Reading the Mind in the Eyes} was originally developed by \cite{baron1997another} and was revised in 2001 \cite{baron2001reading}. In general, the task was designed to detect subtle deficits in emotional recognition. It has been used in a variety of domains, including brain studies \cite{adolphs2002impaired}, dementia \cite{gregory2002theory}, and clinical disorders (e.g., \cite{fett2011relationship}). It has even been shown to correlate with collective intelligence and social sensitivity \cite{woolley2010evidence}. Recently, a brief, 10-item version of this test was proposed by \cite{olderbak2015psychometric}. This version is significantly shorter but retains the diagnostic value of the original. Therefore, it was seen as a superior option in this context. Although it measures emotion recognition rather than mindreading per se (see \cite{oakley2016theory}), it remains one of the most widely used assessment tools for mental state reasoning. The participants were shown 10 images, one at a time, and asked which of four potential emotions was exhibited in each. The maximum score was 10 and the minimum was 0.

The \textit{Interpersonal Reactivity Index} (IRI) is currently one of the most widely used tests for various forms of empathy \cite{davis1980multidimensional}. Empathy is commonly linked with ToM, as those who are better able to experience the emotions of others should be better able to understand others' perspectives. Although IRI is not a measure of ToM per se, IRI appears to be predictive of performance on other tasks of social cognition \cite{eddy2019you}. Each question contained an empathetic statement. Participants answered how well each statement described them on a 5-point Likert scale, with a range from ``Does not describe me well'' (0) to ``Describes me well'' (4). The maximum score could be 28 on each subscale or 112 in general. The minimum would be 0.

\subsubsection{Spatial Reasoning}

The \textit{Perspective Taking Questionnaire} was inspired by \cite{hegarty2004dissociation} and used by \cite{nguyen2022theory}. It involved four questions; each presenting eight icons (stop sign, car, etc.) positioned in a circle. Participants were allowed to study the grid for as long as they wanted. They were also given a hint about what item the question would involve. After continuing, participants were asked to imagine a person standing at a particular icon in the grid and then to identify what direction they would need to look in order to point to a third object. The maximum score was 4 and the minimum was 0.

The \textit{Spatial Visualizations Questionnaire} was inspired by \cite{burte2019knowing}. In it, participants were shown an unfolded cube and asked to choose which of our cubes could not be formed given the cutout. Participants were asked 10 questions. The maximum score was 10 and the minimum was 0. Although this task did not specifically involve taking the perspective of another person, this type of task is commonly used to assess spatial reasoning abilities.








\subsection{Rock-Paper-Scissors Task} \label{subsubsect:rps}

RPS is a two-player game that has been used by researchers to study competitive behavior in many naturalistic environments \cite{fisher2008rock}. In the first two experiments, participants were explicitly informed that they would be matched with a bot. In the third experiment, participants were explicitly told that they would be connected to another human player. Detailed instructions for the RPS task can be found in Appendix~\ref{appendx:tom-survey}. Before starting the human RPS task, participants entered a ``waiting room'' until they could be grouped into pairs. In each experiment, we expected that participants with higher ToM scores would be able to better predict the bot's actions in RPS, which would result in higher scores.

The RPS game consisted of 20 trials. After each trial, participants were shown the opponent's choice and the number of points earned. 2 points were earned for a WIN (P>R, S>P, R>S), 0 points for a LOSS, and 1 point for a tie. An example of feedback after a tie is shown in Figure \ref{fig:study-design}.

The main difference between the three studies was the type of RPS player that played with the human participant. In experiment 1, participants played against a static-strategy bot that followed a 100\% Win-Stay, Loose-Upgrade strategy. For example, if the bot loses with Scissors, it would always select Rock as its next choice. If it lost with Rock, it would always pick Paper. This simplified strategy should make the bot's actions easier for participants to predict.

In Experiment 2, participants played against a stochastic-strategic bot. The bot followed a Win-Stay, Loose-Upgrade (with 0.8 probability), and Loose-Downgrade (with 0.2 probability) strategy. For instance, if the bot lost with Scissors, it would select Rock with a 0.8 probability and Paper with a 0.2 probability. Similarly, if the bot lost with Rock, it would choose Paper with a 0.8 probability and Scissors with a 0.2 probability. This introduced stochasticity, making it more challenging for participants to accurately predict the bot opponent's moves.

In experiment 3, participants played against a human opponent. Therefore, the participants' actions completely determined RPS outcomes. This experiment aimed to examine how ToM abilities were activated differently when facing a real human opponent compared to a computer-programmed bot.


\subsubsection{Study 1: Static-strategy Bot Opponent}

In this study, participants first completed the RPS block and then the ToM Survey. 87 participants completed the study (age: 40.2 $\pm$ 9.4; 42 female; 73 have a university degree or higher).

\subsubsection{Study 2: Dynamic-strategy Bot Opponent}

In this study, the order of the RPS and ToM Survey was randomly assigned. 96 participants fully completed the study (age: 39.3 $\pm$ 11.1; 39 female, 1 non-binary; 83 have a college degree or higher). About half of the participants completed the RPS block first and then the ToM Survey block (n=54), while the other half completed the ToM Survey block first and then the RPS block (n=42).

\subsubsection{Study 3: Human Player Opponent}
Participants played the RPS block and then the ToM survey. In total, we obtained the total data of 250 participants, 125 pairs (age: 39.7 $\pm$ 10.8; 113 female, 1 non-binary; 230 have a college degree or higher). 

\subsection{Depended Variables} \label{subsec:mesurement}
We collect the following objective performance measures for our hypotheses:
\begin{itemize}
    \item \textbf{Theory of Mind (ToM) metrics}: Given that each ToM questionnaire had a different scale, we standardized the data by computing the z-score for each metric as follows:
    \begin{equation} \centering
        M_{i[z-score]} = \frac{(M_i - \mu_i)}{\sigma_i},
    \end{equation}
    where $M_{i[z-score]}$ represents the standardized value of metric $M_i$, $\mu_i$ is the mean of metric $M_i$ across all human subjects who completed the ToM survey in the three studies, and $\sigma_i$ is the standard deviation of metric $M_i$ for all subjects.

    \item \textbf{Effectiveness}: We define the effectiveness of human decision-making in the RPS game as the percentage score each player earns over multiple rounds. More precisely, it is calculated by dividing the total earned score by the maximum possible score across 20 game rounds.
    
    \item \textbf{Prediction Accuracy}: Prediction accuracy is defined as how accurately each player can predict the opponent's choice. This metric is calculated by determining the number of accurate predictions in which the player's choice (assuming they always want to win) would defeat the opponent's actual choice in each round of the game. For instance, if the player chooses `rock', it means they predict their opponent will choose `scissors'. The metric calculates the ratio of accurate predictions to the total number of rounds played, providing a measure of the player's predictive success.

\end{itemize}

\section{Results}
\label{sec:simulation}
In total, we collected data from 433 participants who were able to complete all tasks in our three studies. In the following, we answer our research questions based on the data collected from these participants.

\subsection{RQ1: How are the Theory of Mind (ToM) metrics correlated with each other?} \label{subsec:rq1}
We start by examining the interaction between different cognitive, spatial reasoning, and social-emotional perceptiveness metrics in the three studies. We standardized these metrics following the methodology described in Section \ref{subsec:mesurement} and then calculated the Pearson correlation to analyze the relationships between these standardized ToM variables.

\begin{figure*}[!htpb]
\vspace{-1ex}
\begin{center}
\includegraphics[width=.9\linewidth]{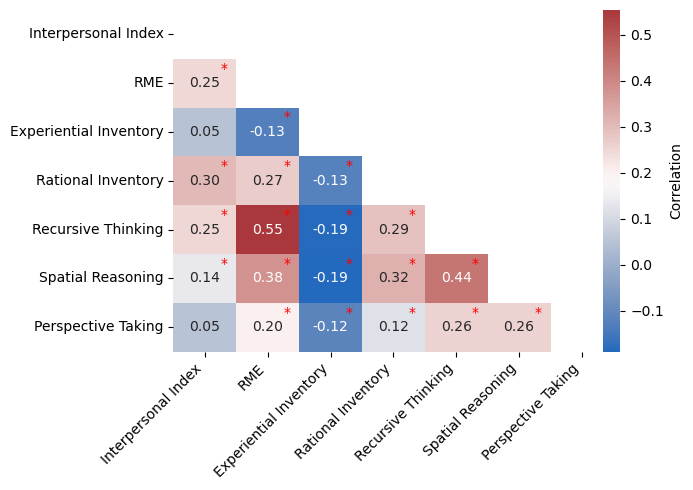} 
\caption{Pearson correlations between ToM metrics across the three studies. Significant correlations, with $p < 0.05$, are marked with an asterisk (*).
\label{fig:corr}
} 
\end{center}
\vspace{-1ex}
\end{figure*}

Figure~\ref{fig:corr} displays the Pearson correlation matrix among the standardized cognitive, spatial reasoning, and social-emotional metrics. The results indicate the strongest correlation ($r=0.55$, $p < 0.0001$) between ``recursive thinking'' and ``RME'' (Reading the Mind in the Eyes), suggesting that there is a strong relationship between the ability to think recursively and the interpretation of mental states from facial expressions.

We also found significant strong correlations, including ``spatial reasoning'' with both ``recursive thinking'' ($r=0.44$, $p < 0.0001$) and ``RME'' ($r=0.38$, $p < 0.0001$), which indicate a relationship between cognitive reasoning, social-emotional perceptiveness, and spatial reasoning ability. In contrast, the ``experiential inventory'' consistently shows negative correlations with ``RME'' ($r=-0.13$), cognitive metrics (e.g., $r=-0.19$ with ``recursive thinking'' and $r=-0.13$ with "rational inventory"), and spatial awareness metrics (e.g., $r=-0.19$ with ``spatial reasoning'' and $r=-0.12$ with ``perspective taking''). The results suggest that more intuitive processing styles may diverge from structured cognitive and spatial logic tasks.

Interestingly, the ``interpersonal reactivity index'' a measure of empathy, demonstrates moderate correlations with the ``rational inventory'' ($r=0.30$, $p < 0.0001$), ``recursive thinking'' ($r=0.249$, $p < 0.0001$) and RME ($r=0.246$, $p < 0.0001$). These results somehow highlight the connection between interpersonal empathy and cognitive skills related to social perceptiveness.


In summary, the strong correlations between cognitive and social-emotional metrics, along with the significant association with spatial reasoning, indicate the interplay among these factors rather than their isolation. These findings corroborate previous neuroimaging studies showing overlap and specialization in brain functioning for ToM tasks associated with these three factors \cite{schurz2014fractionating}. This interplay suggests the need for an integrated approach in assessing ToM and understanding how these factors collectively influence human decision making.

\subsection{RQ2: How do human players adjust their strategies when competing against bots versus other humans in a repetitive adversarial game?} \label{subsec:rq2}

\begin{figure}[ht!]
    \centering
    \begin{subfigure}[b]{0.49\textwidth}
        \includegraphics[width=\textwidth]{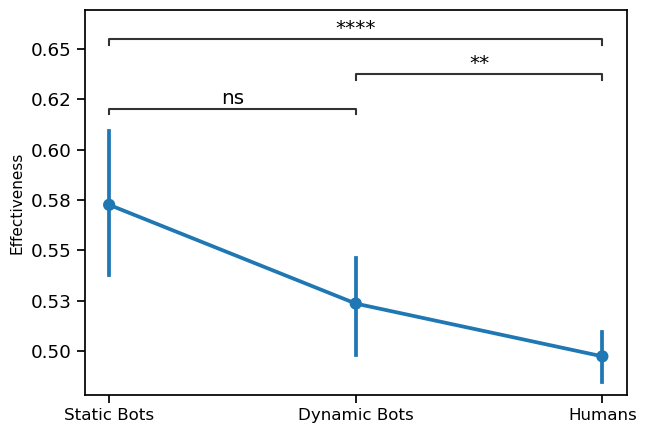}
        \caption{\textbf{Effectiveness}}
        \label{fig:percent-score}
    \end{subfigure}
    \hfill
    \begin{subfigure}[b]{0.49\textwidth}
        \includegraphics[width=\textwidth]{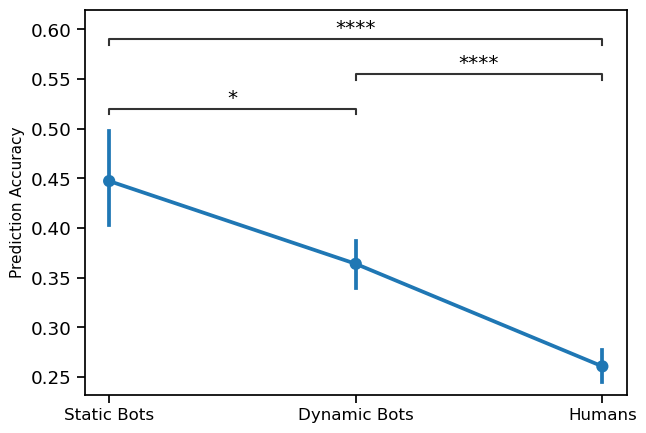}
        \caption{\textbf{Prediction accuracy}}
        \label{fig:pred-accuracy}
    \end{subfigure}
    \caption{RPS performance of participants in terms of (a) effectiveness and (b) prediction accuracy when playing with static bots, dynamic bots, and humans. Error bars indicate 95\% confidence intervals. Note that the two plots have different ranges on the $y-$axis with percent score in $[0.5, 0.65]$ and prediction accuracy in $[0.25, 0.6]$}
    \label{fig:rps}
\end{figure}

Next, we investigate how humans adapt to different types of opponents in the RPS game. 
We analyze the effects of these types of opponents on the human player's performance in terms of effectiveness and prediction accuracy in the RPS game when paired with static bots, dynamic bots, and humans.

Figure~\ref{fig:percent-score} displays the result in comparing the effectiveness of human players against different opponents. Statistical analysis using the Mann-Whitney-Wilcoxon test with a two-sided approach and Bonferroni correction for multiple comparisons shows that human players performed significantly better when facing static bots compared to when playing against stochastic bots ($p < 0.01$) and other human players ($p < 0.0001$). We did not find a significant difference in effectiveness when humans were paired with bots using the stochastic strategy and when paired with those using the static strategy ($p > 0.05$). These results highlight the varying levels of adaptability in human decision-making and the ability to exploit opponents' patterns when confronted with different types of opponents in repeated adversarial scenarios.
 
Comparison of prediction accuracy between different types of opponents also shows statistically significant differences (Figure \ref{fig:pred-accuracy}). We found that when human players interacted with static bots, their prediction accuracy was significantly higher compared to interactions with stochastic bots ($p=0.02$) and encounters with other human players ($p < 0.0001$). Similarly, prediction accuracy against stochastic bots was significantly higher compared to playing against human opponents ($p < 0.001$). These findings underscore the human ability to recognize exploitable patterns and the complexity of predicting opponents with dynamic strategies in competitive settings. This leads us to question how this relates to their ToM ability, which is captured by the three factors: cognitive abilities, spatial reasoning, and social-emotional perceptiveness, prompting our next research question.

\subsection{RQ3: How does measuring human Theory of Mind (ToM) ability, assessed through ToM metrics, impact their performance in repetitive RPS game?} \label{subsec:rq3}

\begin{figure}[!htbp]
\centering
\begin{subfigure}[b]{1\linewidth}
        \caption{\textbf{Static Bots}}
       \begin{subfigure}[b]{0.49\textwidth}
        \includegraphics[width=\textwidth]{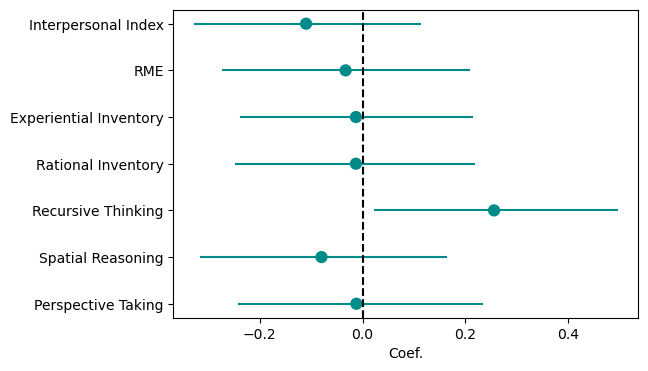}
        \caption*{Effectiveness}
        \label{fig:coef-percent-score-static}
    \end{subfigure}
    \hfill
    \begin{subfigure}[b]{0.49\textwidth}
        \includegraphics[width=\textwidth]{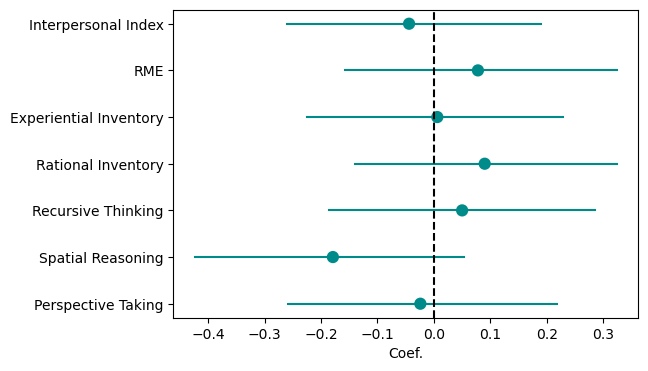}
        \caption*{Prediction accuracy}
        \label{fig:coef-pred-accuracy-static}
    \end{subfigure}
\label{fig:exp2-static-rps}
    \end{subfigure} 
 \begin{subfigure}[b]{1\linewidth}
     \caption{\textbf{Dynamic Bots}}
        \begin{subfigure}[b]{0.49\textwidth}
        \includegraphics[width=\textwidth]{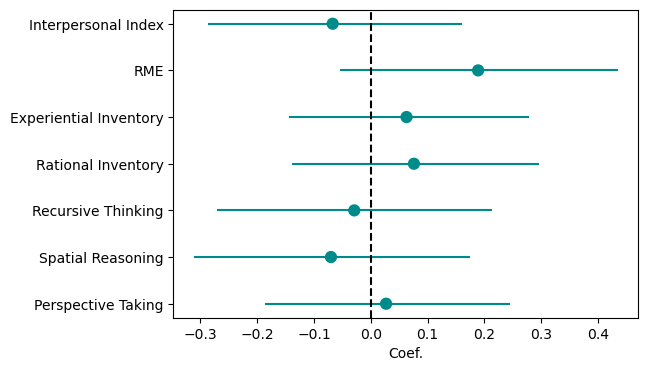}
        \caption*{Effectiveness}
        \label{fig:coef-percent-score-stochastic}
    \end{subfigure}
    \hfill
    \begin{subfigure}[b]{0.49\textwidth}
        \includegraphics[width=\textwidth]{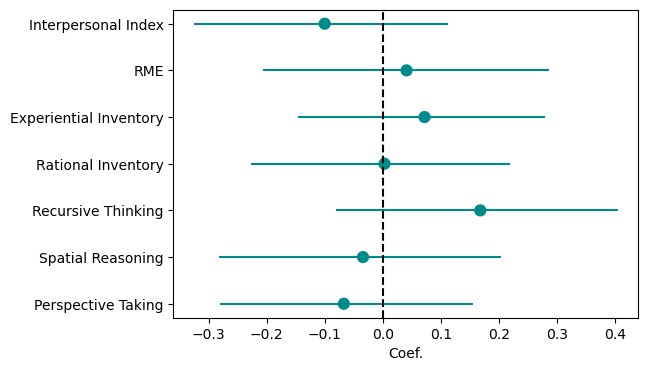}
        \caption*{Prediction accuracy}
        \label{fig:coef-pred-accuracy-stochastic}
    \end{subfigure}
\label{fig:exp1-stochastic-rps}
    \end{subfigure}\hspace{1mm} 

\begin{subfigure}[b]{1\linewidth}
    \caption{\textbf{Human Players}}
       \begin{subfigure}[b]{0.49\textwidth}
        \includegraphics[width=\textwidth]{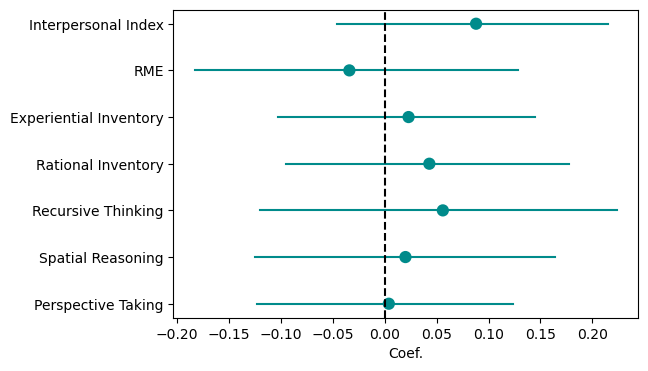}
        \caption*{Effectiveness}
        \label{fig:coef-percent-score-human}
    \end{subfigure}
    \hfill
    \begin{subfigure}[b]{0.49\textwidth}
        \includegraphics[width=\textwidth]{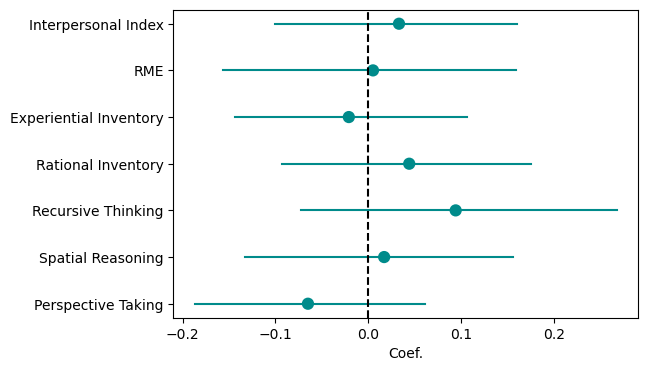}
        \caption*{Prediction accuracy}
        \label{fig:coef-pred-accuracy-human}
    \end{subfigure}
\label{fig:exp3-human-rps}
    \end{subfigure} 
\caption{Estimated coefficients from linear regression models for predicting the player's effectiveness and prediction accuracy when facing static bots, dynamic bots, and human players. The error bars indicate the 95\% confidence intervals.}
\label{fig:rps-corr}
\end{figure}

We examine whether standardized scores on ToM variables regarding cognitive, spatial reasoning, and socio-emotional factors can predict human decision performance in RPS tasks. To this end, we fitted linear regression models to predict human effectiveness and prediction accuracy. Concretely, the model includes seven predictors: interpersonal reactivity index, RME (Reading the Mind in the Eyes), experiential inventory, rational inventory, recursive thinking, spatial reasoning, and perspective taking. The estimated coefficients for various ToM variables, along with their 95\% confidence intervals, are reported in Figure~\ref{fig:rps-corr}. The details of fitting the model's parameters are in Appendix~\ref{appendix:results}.

For static bots (Figure \ref{fig:exp2-static-rps}, left), recursive thinking ability significantly improved decision effectiveness ($\beta=0.26$, 95\% CI = $[0.023, 0.491]$), while other variables did not have a significant impact on effectiveness. The results in terms of prediction accuracy (Figure \ref{fig:exp2-static-rps}, right) suggest that none of the predictor variables significantly affects prediction ability.

The results for players' effectiveness (Figure \ref{fig:exp1-stochastic-rps}, left) against dynamic bots indicate that none of the predictor variables significantly affect the outcome. However, RME shows a marginally significant positive association with decision effectiveness ($\beta=0.19$, 95\% CI = $[-0.056, 0.437]$). In terms of prediction accuracy (Figure \ref{fig:exp1-stochastic-rps}, right), the recursive reasoning ability shows a positive correlation ($\beta=0.17$, 95\% CI = $[-0.075, 0.409]$), but its impact is only marginally significant as the 95\% interval includes zero.

When playing against other human players, we found that most of the ToM variables were insignificant, but the interpersonal variable was positively associated with the players' effectiveness ($\beta=0.1$, 95\% CI = $[-0.041, 0.222]$). In terms of prediction accuracy (Figure \ref{fig:exp3-human-rps}, right), the recursive thinking variable again demonstrates a moderately positive association ($\beta=0.1$, 95\% CI = $[-0.079, 0.265]$).

In general, the findings suggest that no single ToM measurement consistently predicts performance in the RPS game across all settings. However, individuals with higher recursive thinking ability perform slightly better in decision effectiveness against static bots, while RME has a marginally positive impact against dynamic bots. Interpersonal skills are beneficial when playing against human opponents. Recursive thinking ability generally improves prediction accuracy in most scenarios. Since no single ToM measurement significantly impacts decision performance in RPS settings, our next research question is whether these ToM variables belong to the same construct and whether that construct can affect RPS game performance.

\subsection{RQ4: To what extent different ToM metrics are different terms for the same construct, and to what extent can they affect human decision performance in repeated RPS play?} \label{subsec:rq4}

From our investigation into RQ1 (Section \ref{subsec:rq1}) and RQ3 (Section \ref{subsec:rq3}), we discovered strong relationships among ToM metrics
with no single ToM measurement significantly affecting decision performance in RPS settings. Here, our key questions revolve around understanding how these variables relate to the same constructs, namely latent variables, and whether these latent variables predict human decision performance in the RPS game. To this end, we 
used both exploratory factor analysis (EFA) and Structural Equation Models (SEMs) to assess the unique and shared variance within ToM metrics and explain the connections between the constructs and human decisions.


\begin{table}[!htpb]
\centering
\caption{Factor loadings for the maximum-likelihood
EFA factor solution in the three studies.}
\begin{tabular}{lcc|cc|cc}
\toprule
& \multicolumn{2}{c}{Static Bots} & \multicolumn{2}{c}{Dynamic Bots} & \multicolumn{2}{c}{Human Players} \\
\cmidrule(lr){2-3} \cmidrule(lr){4-5} \cmidrule(lr){6-7}
Measure & Factor 1 & Factor 2 & Factor 1 & Factor 2 & Factor 1 & Factor 2 \\
\midrule
Interpersonal Index & -0.11 &\textbf{ 0.53} & 0.34 & \textbf{0.94} & 0.25 & 0.30 \\
RME & 0.31 & \textbf{0.54} & \textbf{0.65} & 0.11 & \textbf{0.68} & 0.18 \\
Experiential Inventory & -0.44 & -0.10 & -0.13 & 0.22 & -0.23 & -0.09 \\
Rational Inventory & 0.26 & 0.41 & 0.45 & 0.10 & 0.12 & \textbf{0.99} \\
Recursive Thinking & 0.32 & 0.46 & \textbf{0.70} & -0.05 & \textbf{0.88} & 0.21 \\
Spatial Reasoning & \textbf{0.63} & 0.10 & \textbf{0.68} & -0.12 & \textbf{0.50} & 0.31 \\
Perspective Taking & \textbf{0.64} & 0.13 & 0.35 & -0.06 & 0.28 & 0.04 \\
\bottomrule
\end{tabular}
\label{tab:factor-loading}
\end{table}

We performed an EFA to assess how well each ToM measure was associated with different constructs or latent variables. Using maximum likelihood estimation and varimax rotation, we extracted two factors with eigenvalues greater than 1 across all three studies (static bots: 2.36 and 1.24; dynamic bots: 2.43 and 1.2; human players: 2.6 and 1.12). The two factors explained 34\% and 18\% of the variance in static bots, 35\% and 14\% in dynamic bots, and 37\% and 16\% in human players. Table \ref{tab:factor-loading} reports the two-factor maximum-likelihood EFA factor for ToM variables in the three studies. The EFA provides preliminary evidence that two factors are needed to describe our ToM survey metric data. We selected only variables with loadings of 0.5 or higher to represent the two latent constructs.

The results in Table \ref{tab:factor-loading}
indicate the patterns of factor loadings for different ToM measures in the three studies. Against static bots, Factor 1 is strongly associated with spatial reasoning (0.63) and perspective taking (0.64), while Factor 2 is linked to the interpersonal index (0.53) and social perceptiveness RME (0.54), highlighting interpersonal and social-emotional dimensions. When facing dynamic bots, Factor 1 is associated with RME (0.65), recursive thinking ability (0.70), and spatial reasoning (0.68), which indicates an association with cognitive, social-emotional, and spatial reasoning processes, while Factor 2 is strongly associated with interpersonal ability (0.94). In the context of human players, Factor 1 is related to recursive thinking (0.88), RME (0.68), and spatial reasoning (0.5), whereas Factor 2 is dominated by the rational inventory (0.99). These findings suggest that Factor 1 generally captures cognitive, social perceptiveness, and spatial aspects of ToM, while Factor 2 aligns more with the interpersonal and rational dimensions.


\begin{figure}[!htbp]
\centering
\begin{subfigure}[b]{1\linewidth}
        \caption{\textbf{Static Bots}}
       \begin{subfigure}[b]{0.49\textwidth}
        \includegraphics[width=.7\textwidth]{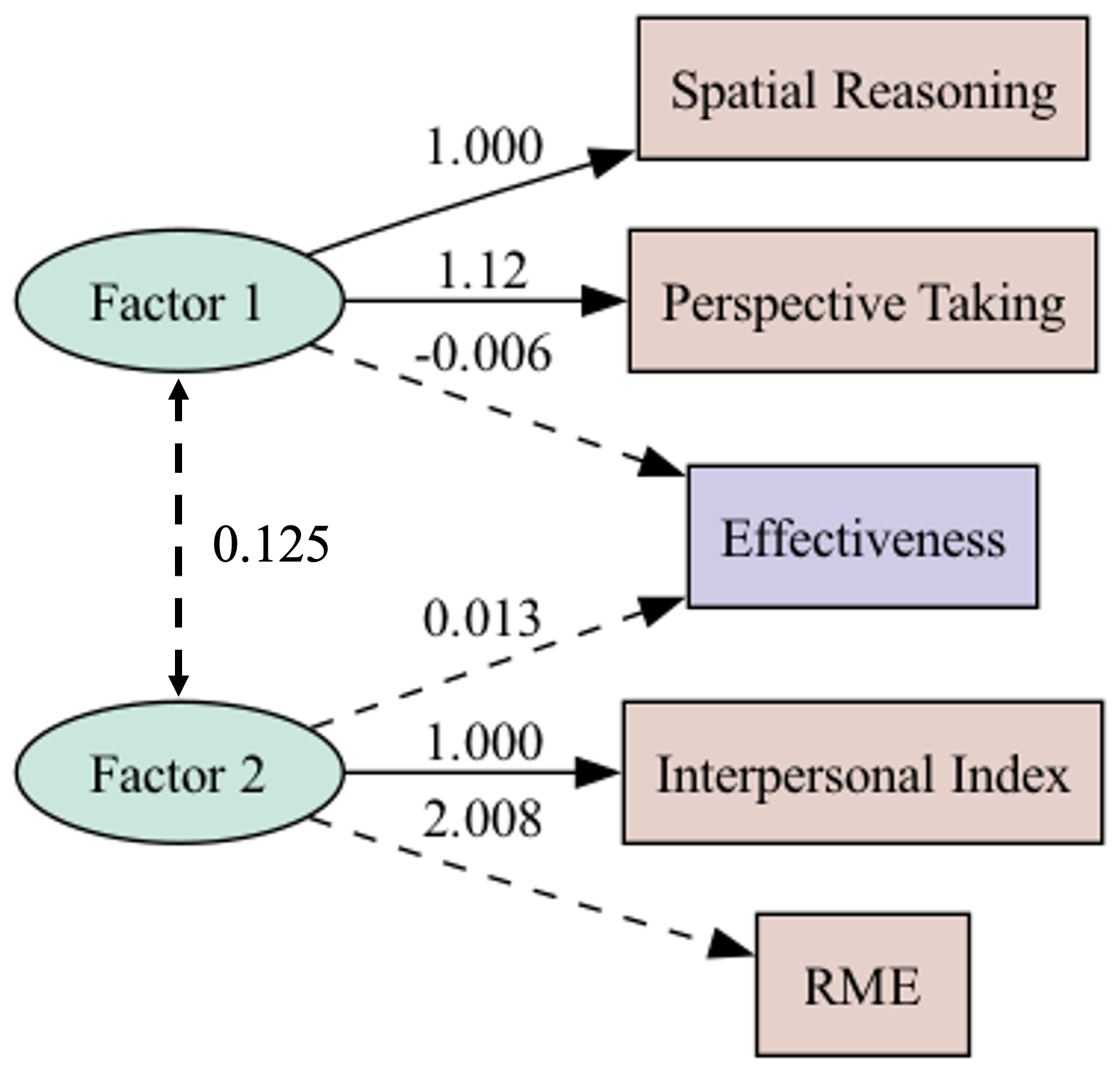}
        \caption*{Effectiveness}
        \label{fig:sem-percent-score-static}
    \end{subfigure}
    \hfill
    \begin{subfigure}[b]{0.49\textwidth}
        \includegraphics[width=.7\textwidth]{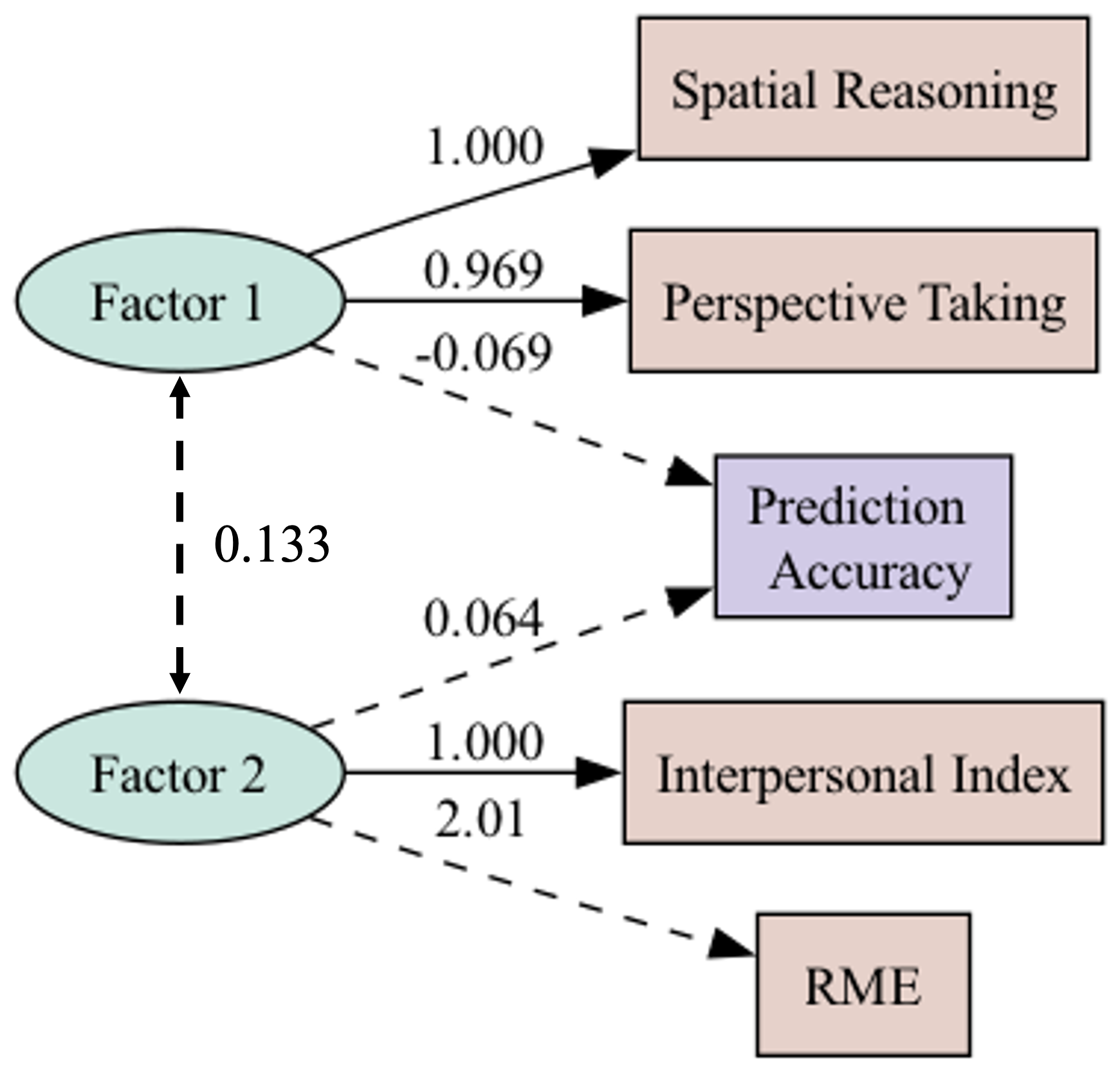}
        \caption*{Prediction accuracy}
        \label{fig:sem-pred-accuracy-static}
    \end{subfigure}
\label{fig:sem-static-rps}
    \end{subfigure} 
 \begin{subfigure}[b]{1\linewidth}
        \caption{\textbf{Dynamic Bots}}
        \begin{subfigure}[b]{0.49\textwidth}
        \includegraphics[width=.7\textwidth]{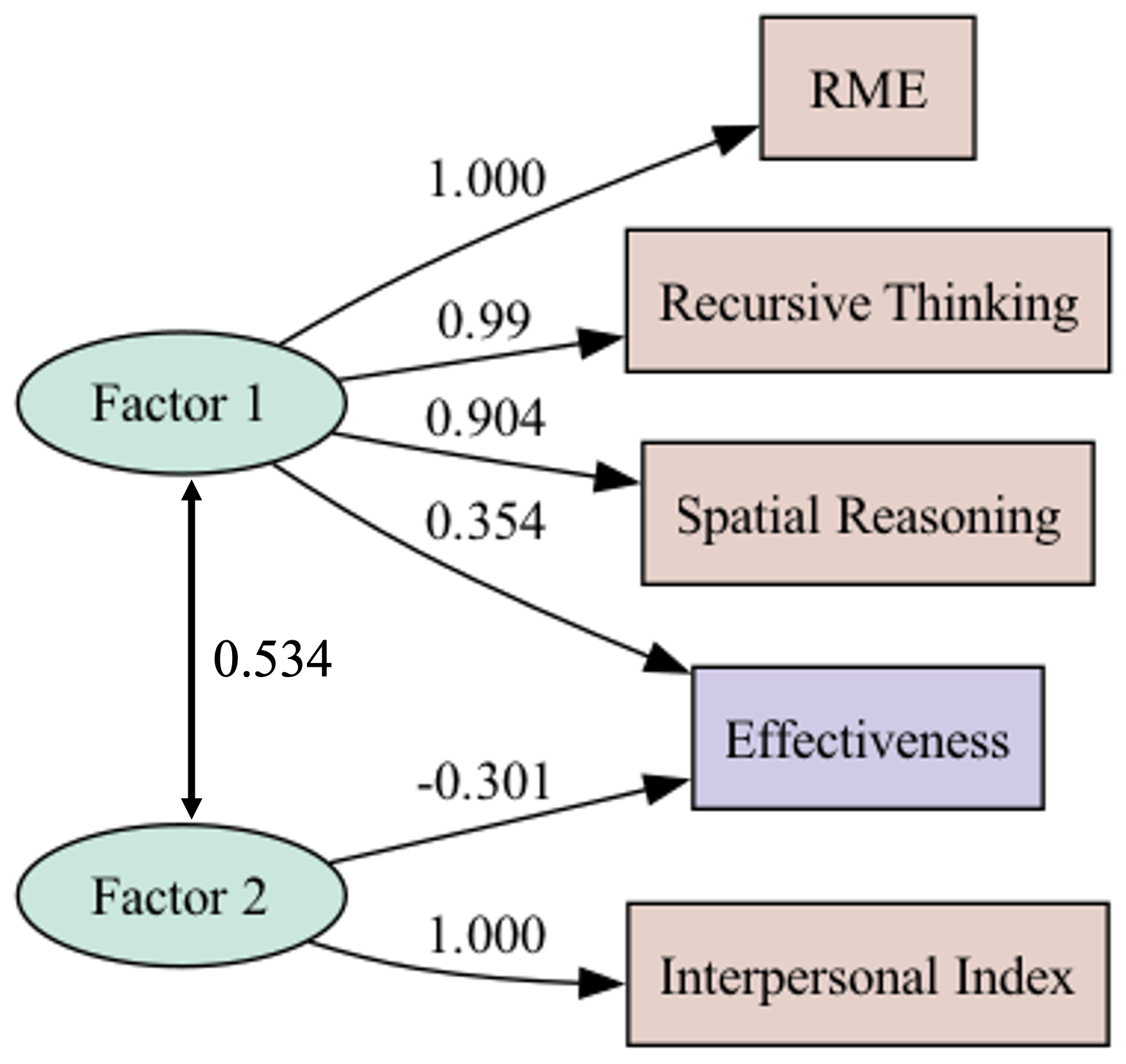}
        \caption*{Effectiveness}
        \label{fig:sem-percent-score-stochastic}
    \end{subfigure}
    \hfill
    \begin{subfigure}[b]{0.49\textwidth}
        \includegraphics[width=.7\textwidth]{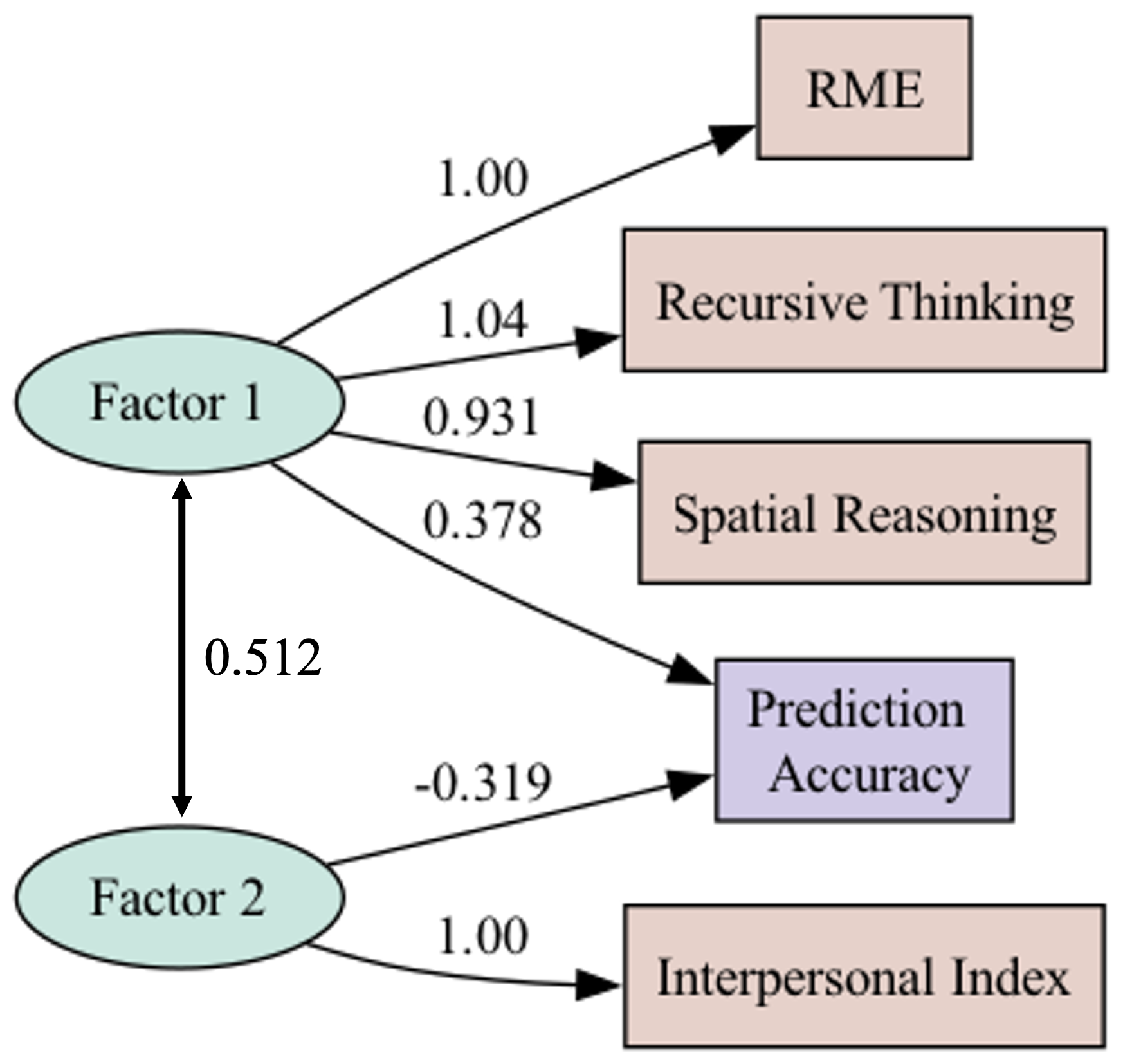}
        \caption*{Prediction accuracy}
        \label{fig:sem-pred-accuracy-stochastic}
    \end{subfigure}
\label{fig:sem-stochastic-rps}
    \end{subfigure}\hspace{1mm} 

\begin{subfigure}[b]{1\linewidth}
        \caption{\textbf{Human Players}}
       \begin{subfigure}[b]{0.49\textwidth}
        \includegraphics[width=.7\textwidth]{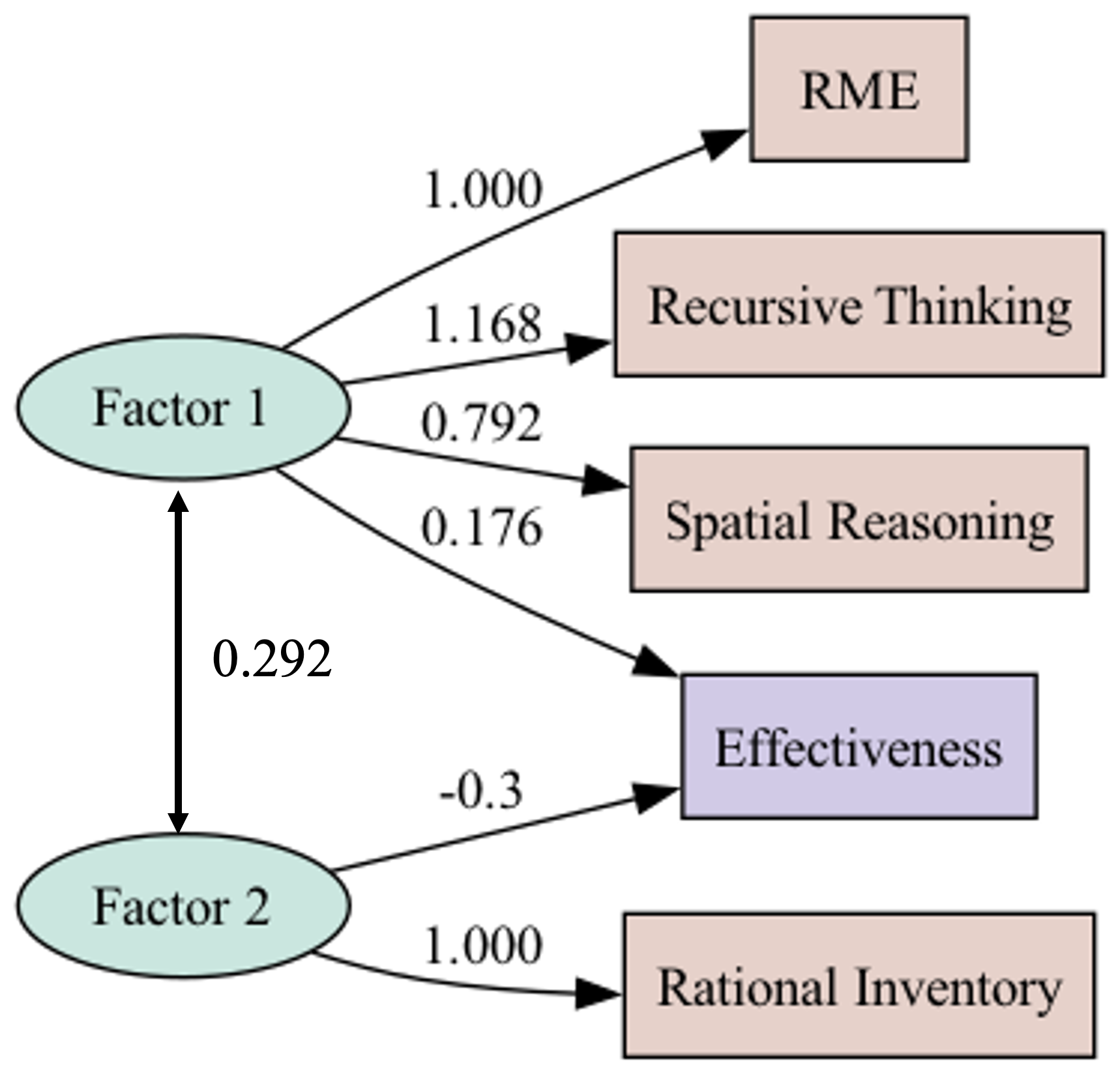}
        \caption*{Effectiveness}
        \label{fig:sem-percent-score-human}
    \end{subfigure}
    \hfill
    \begin{subfigure}[b]{0.49\textwidth}
        \includegraphics[width=.7\textwidth]{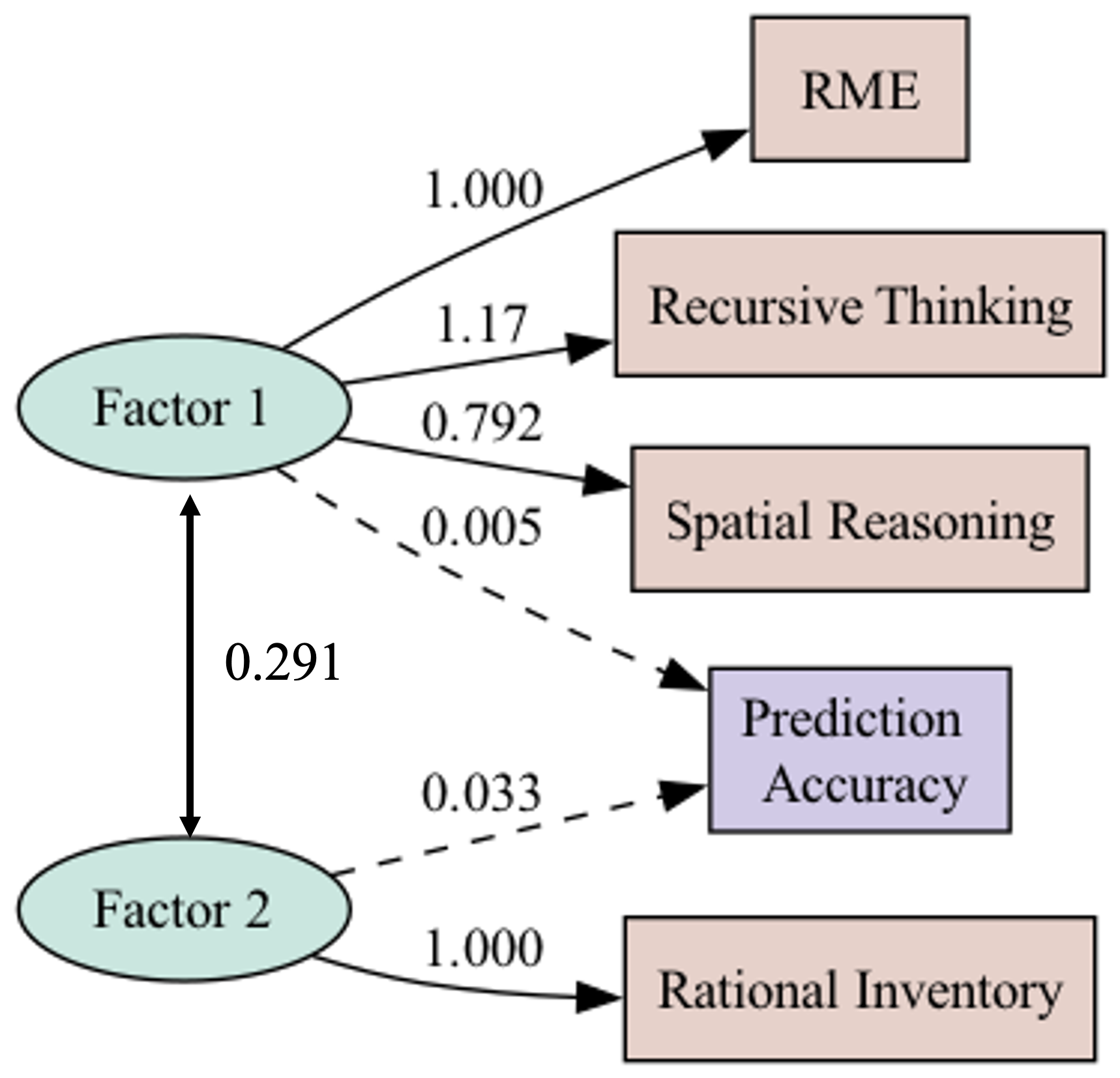}
        \caption*{Prediction accuracy}
        \label{fig:sem-pred-accuracy-human}
    \end{subfigure}
\label{fig:sem-human-rps}
    \end{subfigure} 
\caption{A structural equation model (SEM) for estimating the player's effectiveness and prediction accuracy when facing static bots, dynamic bots, and human players. The numbers on the arrows represent the coefficients. Solid lines indicate paths significant at the 0.05 level.}
 \label{fig:sem}
\end{figure}

Next, we used Structural Equation Models (SEMs) to examine whether the two latent factors relate to human decision-making in the RPS. SEMs enabled us to evaluate the relative contribution of the two latent factors to explaining variation in the individual's effectiveness and prediction accuracy in the RPS game.

\begin{table}[ht]
\centering
\caption{Fit indices for studying the impact of two latent constructs on decision effectiveness in the three studies.}
\begin{tabular}{lcccccc}
\toprule
 & $\chi^2$ Value & $\chi^2$ p-value & CFI & GFI & TLI & RMSEA \\
\midrule
Static Bots & 1.548 & 0.671 & 1.069 & 0.950 & 1.230 & 0.000 \\
Dynamic Bots & 6.010 & 0.111 & 0.947 & 0.910 & 0.824 & 0.103 \\
Human Players & 10.118 & 0.018 & 0.970 & 0.960 & 0.901 & 0.098 \\
\bottomrule
\end{tabular}
\label{tab:fit-indices}
\end{table}

Table~\ref{tab:fit-indices} presents fit indices for studying the impact of two latent constructs on decision effectiveness across three studies. A well-fitting model is indicated by nonsignificant chi-square tests, RMSEA estimates below 0.05, and CFI, GFI, and TLI estimates above 0.90.  The fit indices suggest that the model performs well for static bots, with a nonsignificant $\chi^2$ value (1.548), high CFI (1.069), GFI (0.950), TLI (1.230), and an RMSEA of 0.000. For dynamic bots, the model shows an acceptable fit, with a $\chi^2$ value of 6.010, a $\chi^2$ p-value of 0.111, a CFI of 0.947, a GFI of 0.910, a slightly lower TLI of 0.824, and an RMSEA of 0.103. The model fit for human players has a significant $\chi^2$ value (10.118), a low $\chi^2$ p-value (0.018), but strong CFI (0.970), GFI (0.960), TLI (0.901), and an RMSEA of 0.098.


Figure \ref{fig:sem-static-rps} shows the relationship between the predictor latent variables and the player's effectiveness and prediction accuracy when faced with static bots. Neither Factor 1 nor Factor 2 significantly predict decision effectiveness or prediction accuracy. The results suggest that ToM constructs do not impact these outcomes against easily recognized opponent strategies. Simply put, ToM appears not to be beneficial in this scenario.

 When faced with dynamic bots (Figure \ref{fig:sem-stochastic-rps}), recursive thinking and spatial reasoning significantly impact Factor 1, with estimates of 0.99 and 0.904, respectively, in the model of effectiveness. One loading per factor is fixed to one for model estimation, with RME for Factor 1 and Interpersonal Index for Factor 2. The covariance between Factor 1 and Factor 2 is 0.53 ($p = 0.014$), indicating a significant positive relationship. We found that Factor 1 positively affects effectiveness ($\beta$  = $0.354$, $p < 0.0001$), while Factor 2 negatively affects it ($\beta$  = $-0.301$, $p = 0.0$). Similarly, recursive thinking and spatial reasoning significantly impact Factor 1 in the prediction accuracy model, with estimates of 1.04 and 0.931. The covariance between Factor 1 and Factor 2 is 0.512 ($p = 0.016$). Factor 1 significantly impacts prediction accuracy ($\beta$  = $0.378$, $p = 0.012$), while Factor 2 has a significantly negative impact ($\beta$  = $-0.319$, $p < 0.0001$).

Similarly, against human players, cognitive recursive thinking and spatial reasoning are strongly associated with Factor 1, with estimates of 0.1168 and 0.792. Rational Inventory is used for Factor 2. The covariance between Factor 1 and Factor 2 is 0.292 ($p < 0.001$). Factor 1 positively impacts decision effectiveness ($\beta$ = $0.176$, $p = 0.0$), while Factor 2 has a negative impact ($\beta$  = $-0.3$, $p = 0.0$). In the prediction accuracy model, neither Factor 1 nor Factor 2 had a significant impact.

Overall, the results indicate that ToM constructs do not significantly impact decision effectiveness or prediction accuracy against static bots. However, when facing dynamic bots, recursive thinking, social perceptiveness, and spatial reasoning positively influence effectiveness and prediction accuracy through Factor 1, while Factor 2, characterized by emotional-interpersonal ability, has a negative impact. When playing against human players, Factor 1, characterized by cognitive recursive thinking, social perceptiveness, and spatial reasoning, positively affects decision effectiveness, while Factor 2, associated with rational ability, negatively impacts it; neither Factor 1 nor Factor 2 significantly affects prediction accuracy.

\section{Discussion}
\label{sec:discussion}
As ToM abilities become increasingly important, there is a growing recognition of the need to measure and understand their role in predicting others' actions. Accurately evaluating ToM helps anticipate needs and desires, enabling appropriate responses and decisions. This capability is crucial to improving interactions and collaborations in diverse fields. Studying ToM metrics enables AI systems to model and interpret human behavior, anticipate actions, and respond appropriately, ultimately improving human-machine and human-human collaborations and potentially leading to helpful interventions. Unlike previous work that focuses on one aspect of ToM, this study investigates the interplay among cognitive, emotional, and spatial reasoning, exploring how these interconnected aspects influence human decision behavior and the ability to recognize and exploit opponents' strategies in adversarial settings.

\subsection{Implications for Research and Practice}
Our findings have important implications for behavioral decision science and the modeling of machine ToM in AI systems. The strong correlation between cognitive recursive thinking, spatial reasoning, and social perceptiveness highlights the multifaceted nature of ToM, suggesting that it cannot be fully understood by examining its components in isolation. Developing comprehensive assessments that capture these interrelated aspects is essential. The correlation among these components also indicates that AI models must integrate various cognitive, spatial, and emotional processes to effectively mimic human-like ToM. This integration allows AI systems to better model and interpret human behavior and mental states, resulting in more accurate predictions and interactions. For instance, this initial demonstration could inform the development of machine ToM models in AI and machine learning research. These models can expand beyond using simple gridworld environments for simulating false belief tests \cite{rabinowitz2018machine,nguyen2020cognitive} or intention recognition \cite{baker2017rational,ho2022planning} by incorporating a diverse range of tests that replicate cognitive inquiries, emotional reactions, and spatial relations. This approach is particularly crucial for building sociotechnical systems in collective human-machine intelligence \cite{gupta2023fostering} and to understanding ToM in large language model (LLM) agents, which must be assessed for cognitive, spatial, and emotional aspects, considering both physical and social interactions \cite{shapira2023clever,ullman2023large}.

Our findings have important implications for developing and integrating AI into human interaction. Through exploratory factor analysis (EFA) and structural equation models (SEMs), we identified that cognitive recursive thinking, social perceptiveness, and spatial reasoning positively impact decision effectiveness in dynamic environments, while interpersonal skills negatively affect performance. This suggests that teams can enhance strategic planning and resource allocation by prioritizing these cognitive and social skills for effective decision-making in dynamic contexts ~\cite{almaatouq2024effects}. Additionally, understanding the negative impact of interpersonal skills in dynamic scenarios can inform team composition and training strategies, allowing organizations to optimize performance by selecting and developing the most relevant skills for their teams~\cite{woolley2010evidence,riedl2021quantifying}. Integrating these insights can help organizations foster more adaptive, effective, and strategically aligned teams.

\subsection{Limitations}
The studies we pursued here were initial efforts to shed light on the interplay among the cognitive, emotional, and spatial components of ToM and explore their interconnected aspects in predicting human decision behavior in adversarial settings. This initial exploration revealed interesting insights, but there is considerable room for further investigation. 

While RPS games are beneficial for understanding ToM, they may not fully capture the complexity of real-world social interactions and strategic decision making in adversarial contexts. For instance, many adversarial settings typically involve navigating dynamic environments with delayed feedback, where outcomes are only revealed after a series of decisions, which is not represented in the static rules of RPS. Additionally, real-life situations often involve a mix of cooperation and adversaries in team-based scenarios. Future research should consider more complex and varied scenarios to better understand the role of ToM in these realistic adversarial settings.

In addition, the ToM abilities tests considered in our study may not fully capture the complexity of ToM, potentially missing nuances in how cognitive, emotional, and spatial reasoning contribute in diverse situations. For example, high-order recursive thinking tasks focus on cognitive aspects but overlook false belief tasks. Similarly, tests such as the Reading the Mind in the Eyes (RME) emphasize emotional recognition but do not fully capture empathy or complex social interactions. Future research could benefit from combining different tests to comprehensively capture cognitive, spatial, and emotional dimensions.

In addition, this work only explores simple bot strategies, which do not capture the full spectrum of adaptive strategies players might use. Future research should explore more diverse scenarios that involve opponents modeled by advanced AI bots equipped with machine ToM models \cite{rabinowitz2018machine,nguyen2022theory,baker2017rational} or that use AI agents based on large language models (LLM). This approach would provide a better understanding of how ToM abilities operate when competing against sophisticated opponents, offering insight into the role of ToM in complex and realistic settings.

\section{Conclusion}
\label{sec:conclusion}
In this paper, we conduct experiments to understand how to quantify Theory of Mind (ToM) ability and investigate if these ToM variables can predict human performance in the most well-known repeated adversarial game, Rock-Paper-Scissors (RPS). We find a strong correlation among spatial reasoning, recursive thinking, and the ability to interpret mental states from facial expressions (Reading Minds in the Eyes test - RME). We also observed that human players performed significantly better when facing static bots compared to playing against other human players and stochastic bots. These findings highlight the varying levels of adaptability in human decision-making and the ability to exploit opponents' patterns when confronted with different types of opponents in the RPS game. Furthermore, we learned that while most ToM metrics did not significantly affect performance in the RPS game, individuals with higher levels of recursive thinking ability exhibited slightly better decision-making performance. Finally, by using exploratory factor analysis (EFA) and structural equation models (SEMs), we discovered that Factor 1, characterized by the observed variables related to cognitive recursive thinking, social perceptiveness, and spatial reasoning, positively contributed to decision effectiveness. In contrast, Factor 2, associated with interpersonal skills and emotional-interpersonal abilities, negatively affected decision effectiveness. Our findings suggest that ToM constructs do not impact performance against static bots but play significant roles against dynamic bots and human players, highlighting the need for further research on applying ToM models for effective collaboration in human and human-machine teaming.



\bibliographystyle{ACM-Reference-Format}
\bibliography{chb_bib}

\appendix
\section{Appendix}

\subsection{Instructions: General and RPS} \label{appendx:tom-survey}


There were three minor differences between the instructions in the bot conditions and the human-human conditions. First, in the bot conditions, the instructions stated, ``Today you will complete a series of surveys and play the game Rock-Paper-Scissors (RPS) with a pre-programmed opponent''. Conversely, in the human-human condition, the instructions stated, ``Today you will complete a series of surveys and play the game Rock-Paper-Scissors (RPS) with an MTurk opponent''. Also, the bot condition included the line, ``You and your opponent will make decisions at the same time across multiple rounds''. In contrast, the human-human condition read, ``You and your MTurk opponent will make decisions at the same time across multiple rounds''. The final difference was in the description of payoffs. The bot condition stated, ``Your payoff in a given round will depend on the decisions that both you and your opponent make'' whereas the human-human condition stated, ``Your payoff in a given round will depend on the decisions that both you and your MTurk opponent make''.



\begin{figure}[h!]
    \centering
    \includegraphics[width=1.0\linewidth]{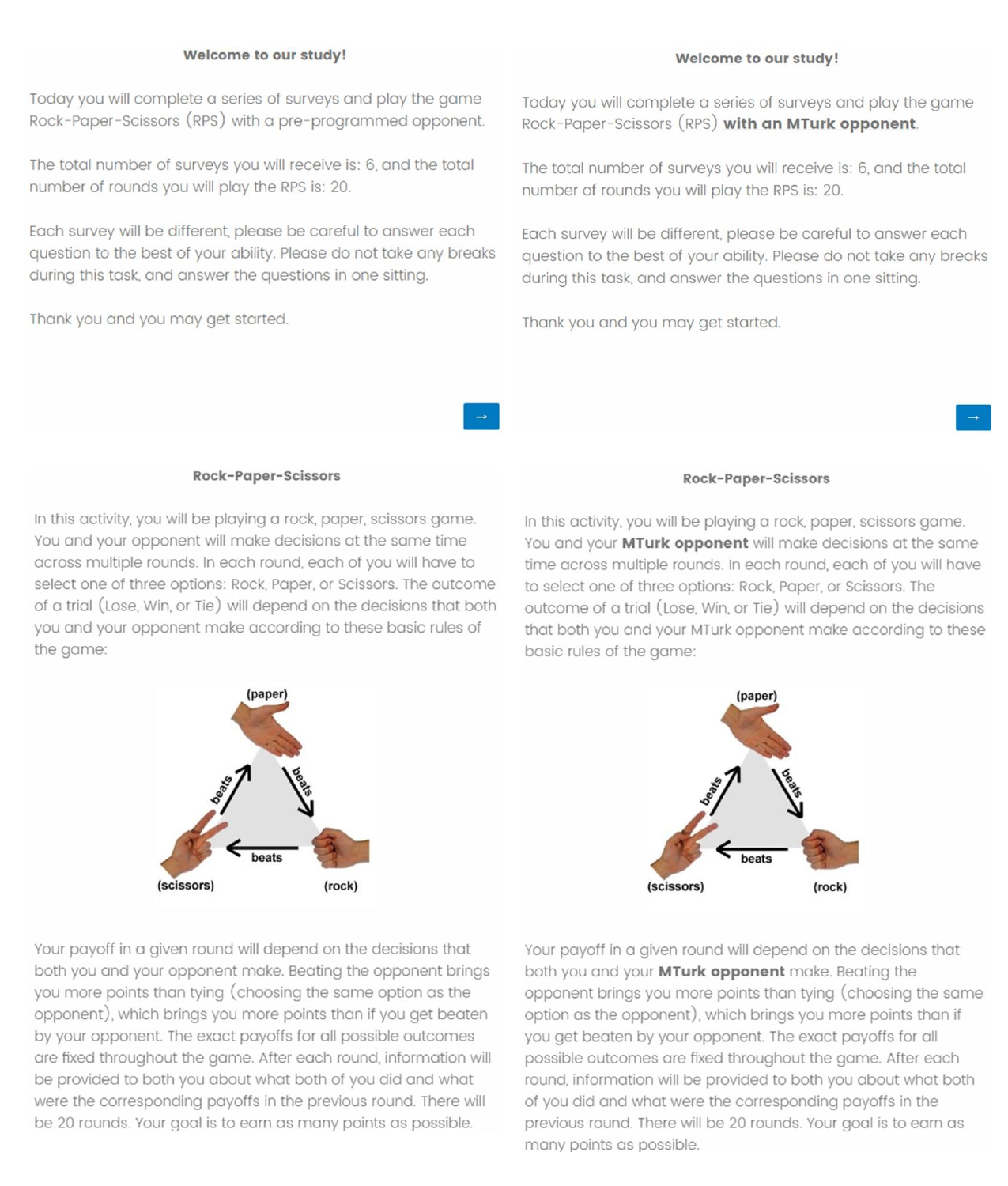}
\end{figure}

\clearpage

\newpage
\subsection{Perspective Taking} \label{subsec:perspectivetaking}

\begin{figure}[h!]
    \centering
    \includegraphics[width=1.0\linewidth]{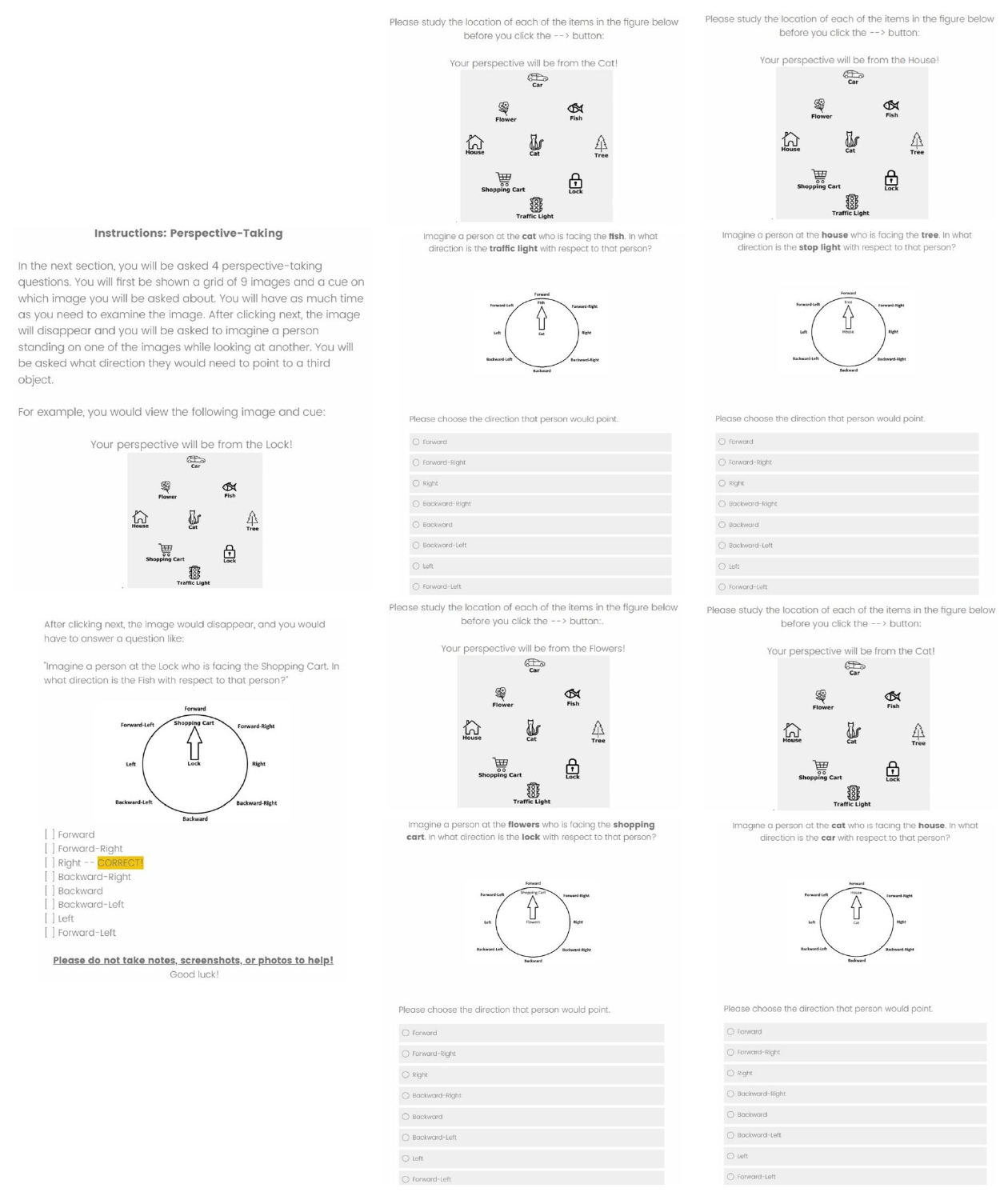}
\end{figure}

\clearpage

\newpage
\subsection{Spatial Rotation} \label{subsec:spatialrotation}

\begin{figure}[h!]
    \centering
    \includegraphics[width=1.0\linewidth]{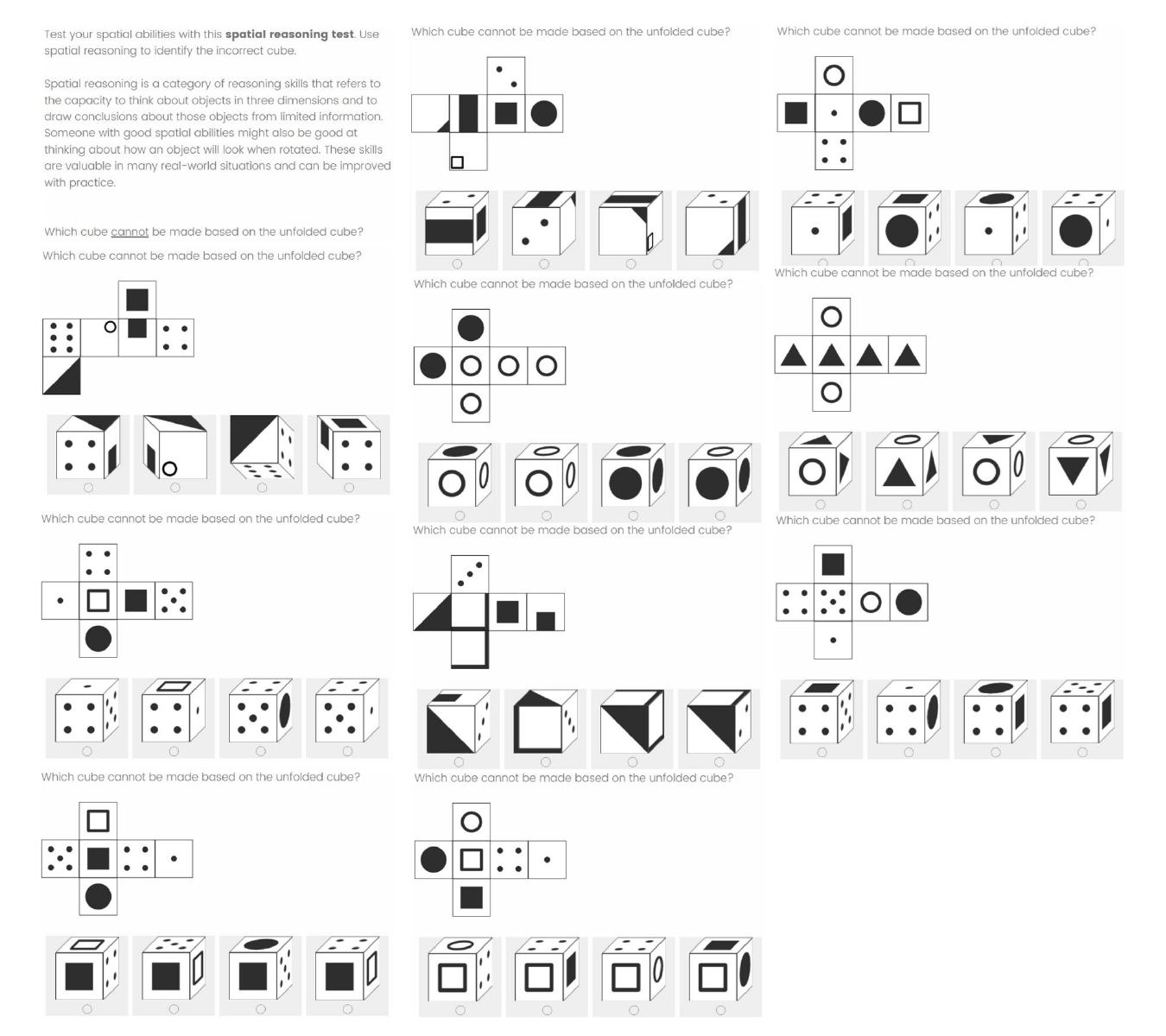}
\end{figure}

\clearpage

\newpage

\subsection{Reading the Mind in the Eyes} \label{subsec:readingeyes}

\begin{figure}[h!]
    \centering
    \includegraphics[width=1.0\linewidth]{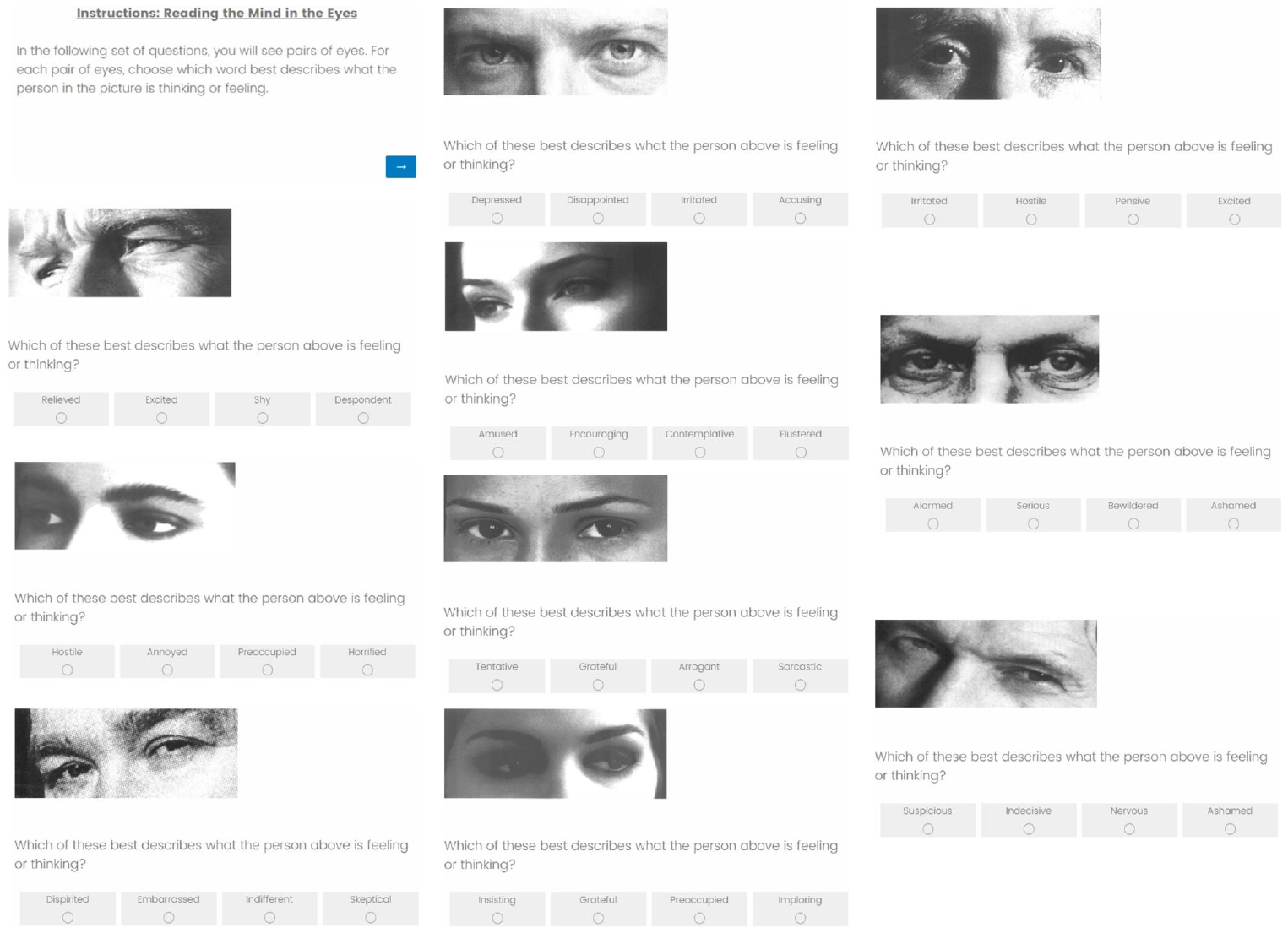}
\end{figure}

\clearpage

\newpage

\subsection{Rational-Experiential Inventory} \label{subsec:readingeyes}

\begin{figure}[h!]
    \centering
    \includegraphics[width=1.0\linewidth]{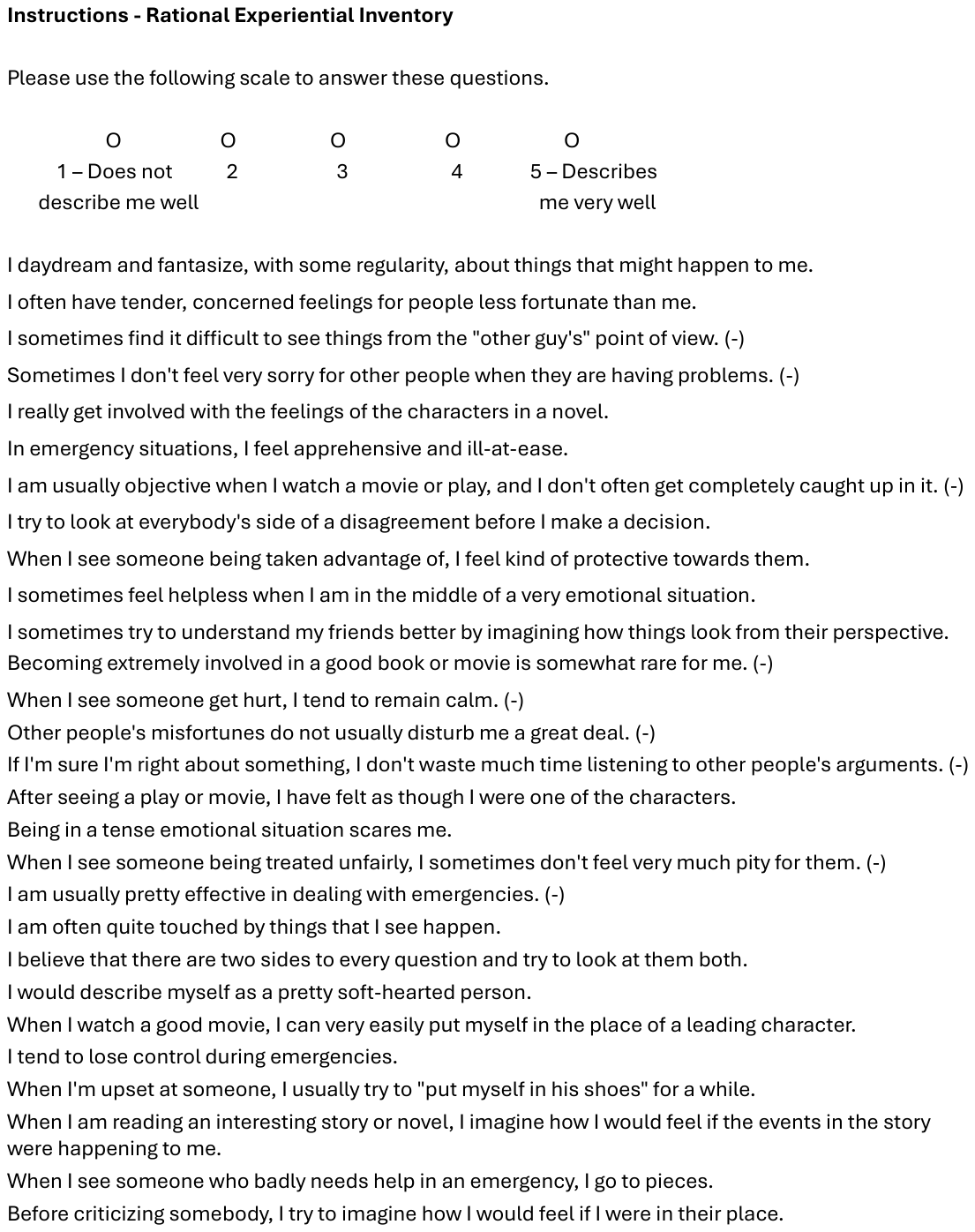}
\end{figure}

\clearpage

\newpage

\subsection{Interpersonal Reactivity Index} \label{subsec:readingeyes}

\begin{figure}[h!]
    \centering
    \includegraphics[width=1.0\linewidth]{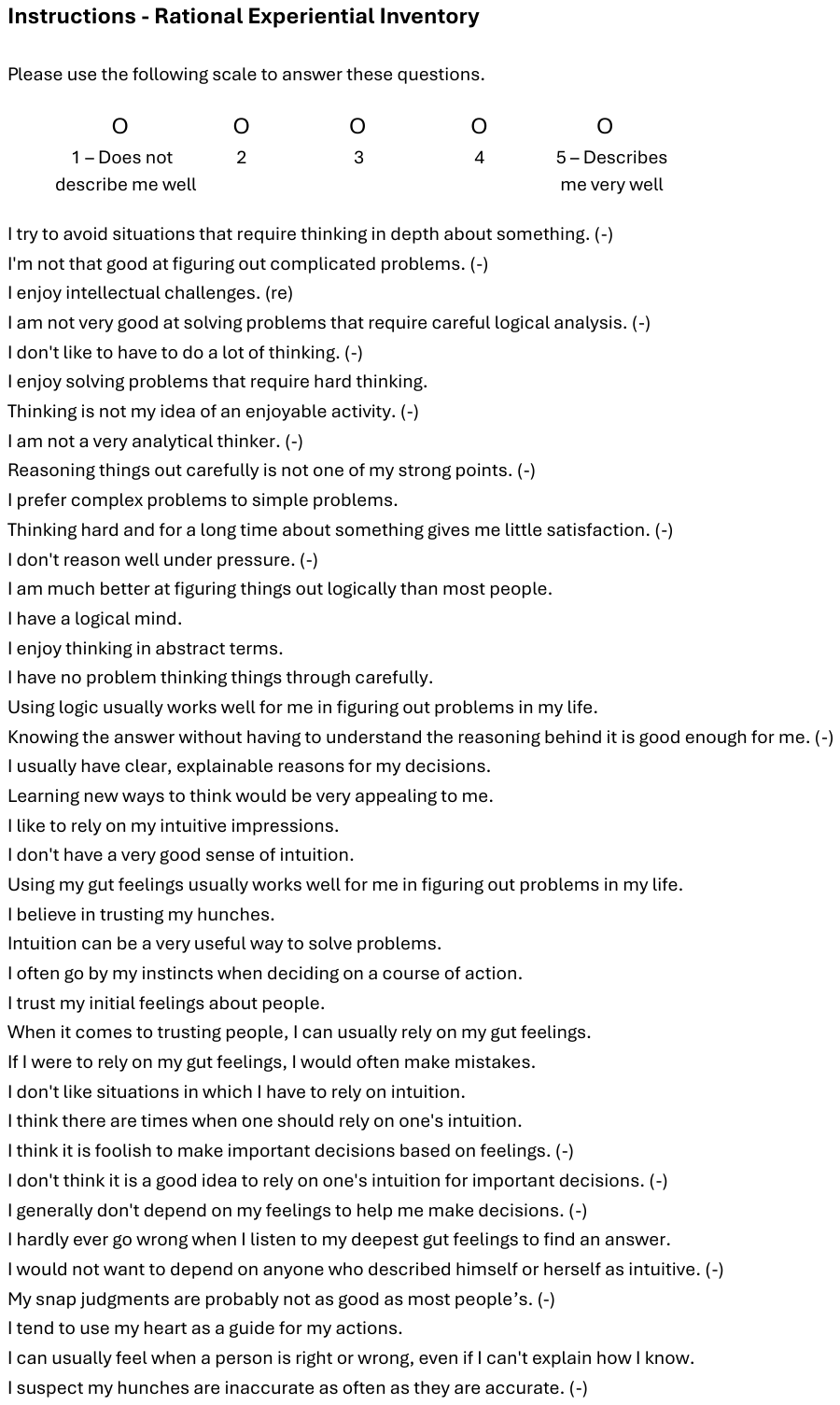}
\end{figure}

\clearpage

\newpage

\subsection{Recursive Thinking Questionnaire} \label{subsec:readingeyes}

\begin{figure}[h!]
    \centering
    \includegraphics[width=1.0\linewidth]{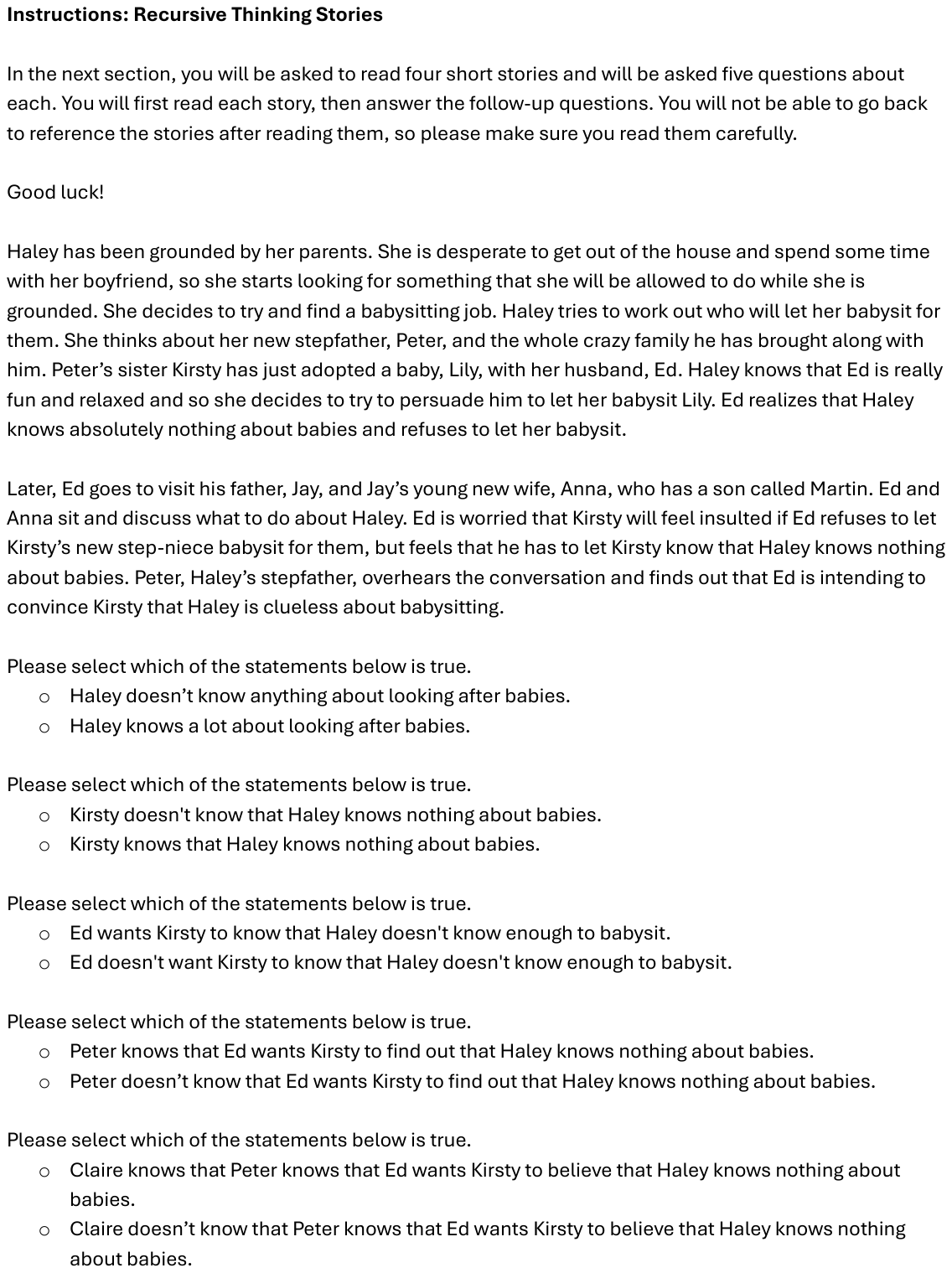}
\end{figure}

\clearpage

\begin{figure}[h!]
    \centering
    \includegraphics[width=1.0\linewidth]{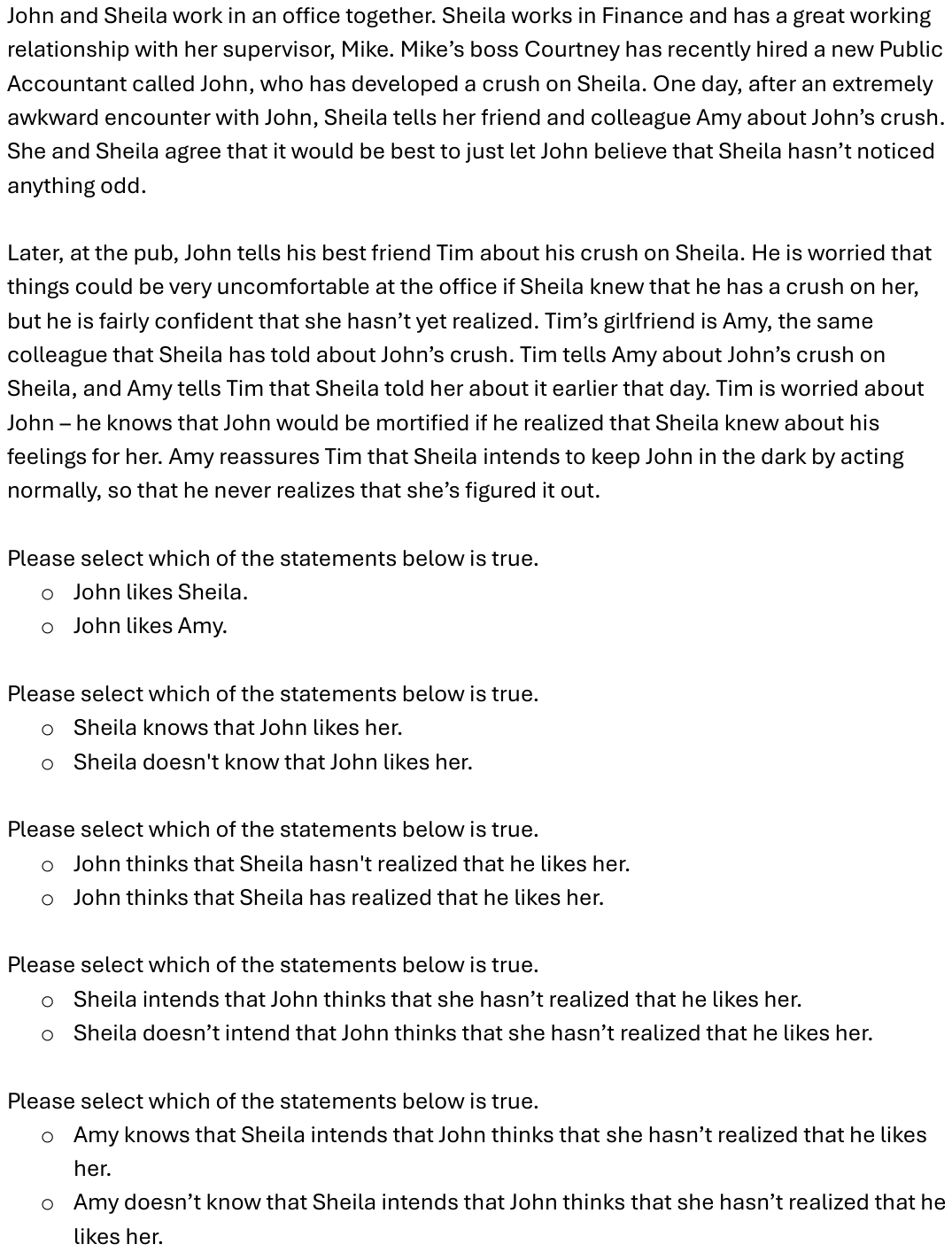}
\end{figure}

\clearpage

\begin{figure}[h!]
    \centering
    \includegraphics[width=1.0\linewidth]{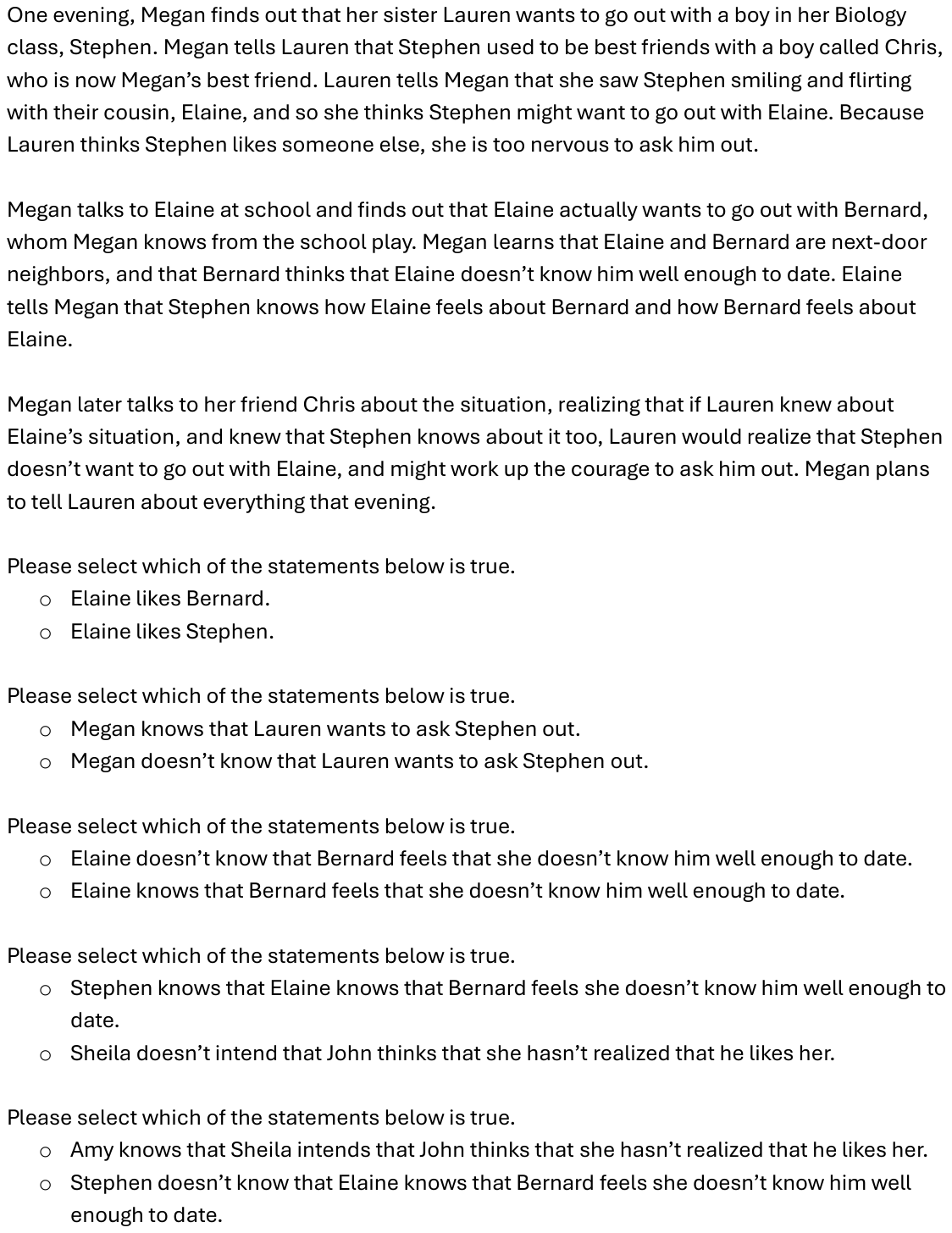}
\end{figure}

\clearpage

\begin{figure}[h!]
    \centering
    \includegraphics[width=0.90\linewidth]{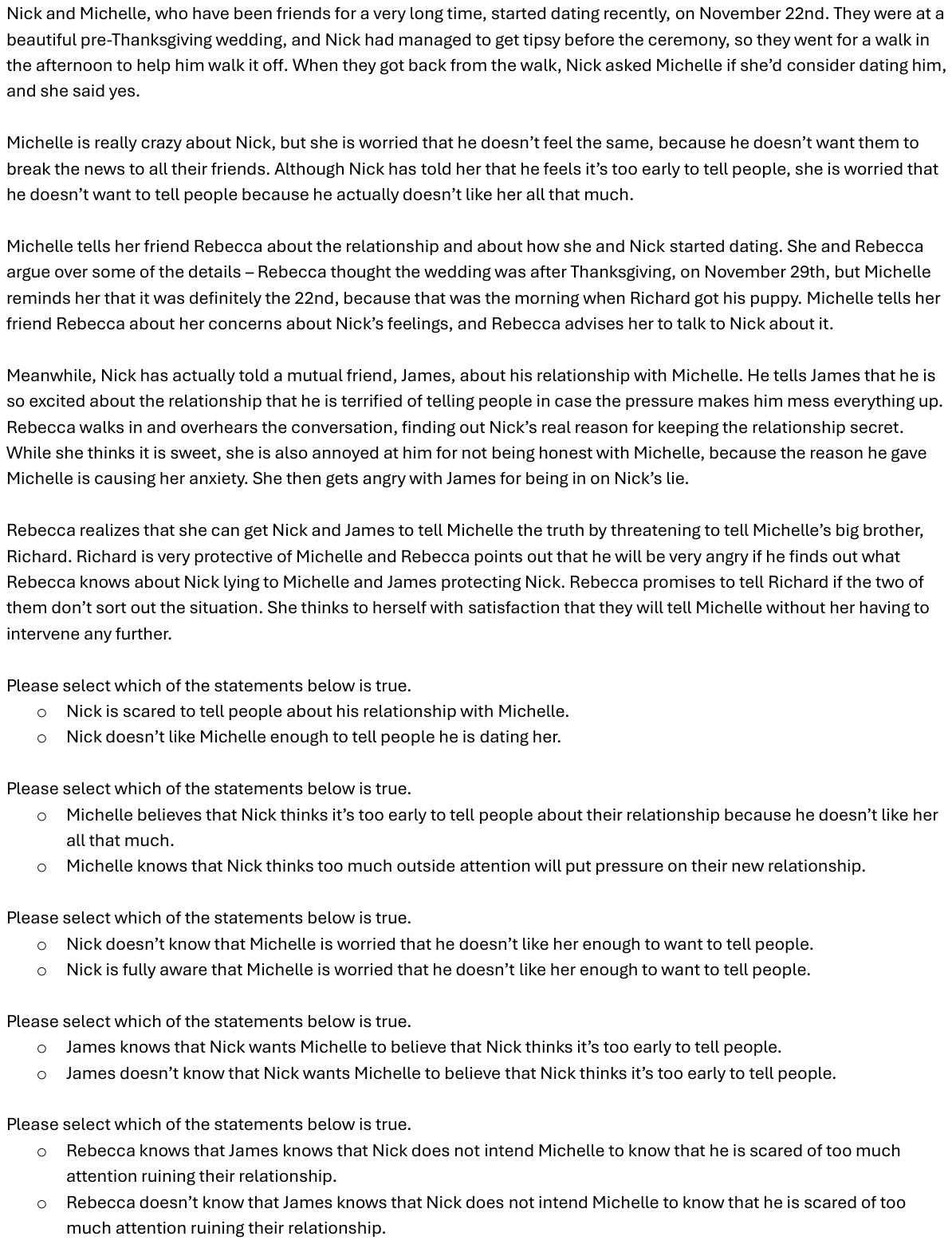}
\end{figure}

\clearpage

\subsection{Results of Fitting Model} \label{appendix:results}

In Section~\ref{subsec:rq3}, we have used a Bayesian linear regression with \texttt{PyMC3}, a Python library for probabilistic programming, to fit models. Here, we present the trace of the sampled values and a summary of the posterior distributions, which includes the mean, standard deviation, and credible intervals, for each model parameter. The coefficients are associated with specific predictor variables: 
\begin{itemize}
    \item $\beta_0$: Interpersonal Reactivity Index
    \item $\beta_1$: Reading the Mind in the Eyes test scores (\textit{RME})
    \item $\beta_2$: Experiential Inventory
    \item $\beta_3$: Rational Inventory
    \item $\beta_4$: Recursive Thinking
    \item $\beta_5$: Spatial Reasoning
    \item $\beta_6$: Perspective Taking
\end{itemize}

\begin{table}[H]
\centering
\caption{Summary of posterior distributions for the parameters in the Bayesian model for estimating the \textbf{decision effectiveness} in the RPS game against \textbf{static bots}.}
\begin{tabular}{lcccccccccc}
\toprule
 & mean & sd & hdi\_3\% & hdi\_97\% & mcse\_mean & mcse\_sd & ess\_bulk & ess\_tail & r\_hat \\
\midrule
alpha & -0.000 & 0.108 & -0.190 & 0.211 & 0.001 & 0.001 & 26550.0 & 15825.0 & 1.0 \\
betas[0] & -0.112 & 0.116 & -0.336 & 0.104 & 0.001 & 0.001 & 24754.0 & 16465.0 & 1.0 \\
betas[1] & -0.034 & 0.128 & -0.271 & 0.214 & 0.001 & 0.001 & 24763.0 & 16428.0 & 1.0 \\
betas[2] & -0.011 & 0.120 & -0.243 & 0.206 & 0.001 & 0.001 & 26385.0 & 16385.0 & 1.0 \\
betas[3] & -0.012 & 0.123 & -0.248 & 0.216 & 0.001 & 0.001 & 24355.0 & 16228.0 & 1.0 \\
betas[4] & 0.258 & 0.125 & 0.023 & 0.491 & 0.001 & 0.001 & 22760.0 & 14952.0 & 1.0 \\
betas[5] & -0.080 & 0.127 & -0.310 & 0.174 & 0.001 & 0.001 & 24270.0 & 15914.0 & 1.0 \\
betas[6] & -0.012 & 0.127 & -0.248 & 0.226 & 0.001 & 0.001 & 25624.0 & 16038.0 & 1.0 \\
sigma & 1.025 & 0.082 & 0.878 & 1.183 & 0.001 & 0.000 & 20682.0 & 13759.0 & 1.0 \\
\bottomrule
\end{tabular}
\label{tab:summary-effectiveness-static}
\end{table}

\begin{table}[H]
\centering
\caption{Summary of posterior distributions for the parameters in the Bayesian model for \textbf{prediction accuracy} when playing against \textbf{static bots}.}
\begin{tabular}{lcccccccccc}
\toprule
 & mean & sd & hdi\_3\% & hdi\_97\% & mcse\_mean & mcse\_sd & ess\_bulk & ess\_tail & r\_hat \\
\midrule
alpha & -0.000 & 0.111 & -0.207 & 0.214 & 0.001 & 0.001 & 27117.0 & 16304.0 & 1.0 \\
betas[0] & -0.044 & 0.120 & -0.266 & 0.185 & 0.001 & 0.001 & 25064.0 & 16571.0 & 1.0 \\
betas[1] & 0.078 & 0.130 & -0.161 & 0.327 & 0.001 & 0.001 & 24536.0 & 15221.0 & 1.0 \\
betas[2] & 0.007 & 0.122 & -0.223 & 0.239 & 0.001 & 0.001 & 26348.0 & 16023.0 & 1.0 \\
betas[3] & 0.092 & 0.124 & -0.134 & 0.333 & 0.001 & 0.001 & 24931.0 & 16248.0 & 1.0 \\
betas[4] & 0.049 & 0.128 & -0.196 & 0.280 & 0.001 & 0.001 & 26089.0 & 16345.0 & 1.0 \\
betas[5] & -0.178 & 0.127 & -0.409 & 0.068 & 0.001 & 0.001 & 24032.0 & 14762.0 & 1.0 \\
betas[6] & -0.025 & 0.128 & -0.268 & 0.211 & 0.001 & 0.001 & 25365.0 & 15757.0 & 1.0 \\
sigma & 1.038 & 0.083 & 0.891 & 1.198 & 0.001 & 0.000 & 20220.0 & 13863.0 & 1.0 \\
\bottomrule
\end{tabular}
\label{tab:summary-pred-accuracy-static}
\end{table}

\begin{table}[H]
\centering
\caption{Summary of posterior distributions for the parameters in the Bayesian model for estimating the \textbf{decision effectiveness} in the RPS game facing \textbf{dynamic bots}.}
\begin{tabular}{lccccccccc}
\toprule
 & mean & sd & hdi\_3\% & hdi\_97\% & mcse\_mean & mcse\_sd & ess\_bulk & ess\_tail & r\_hat \\
\midrule
alpha & 0.001 & 0.104 & -0.199 & 0.194 & 0.001 & 0.001 & 27619.0 & 15523.0 & 1.0 \\
betas[0] & -0.066 & 0.118 & -0.287 & 0.154 & 0.001 & 0.001 & 24388.0 & 16409.0 & 1.0 \\
betas[1] & 0.189 & 0.132 & -0.056 & 0.437 & 0.001 & 0.001 & 23432.0 & 15667.0 & 1.0 \\
betas[2] & 0.063 & 0.112 & -0.150 & 0.269 & 0.001 & 0.001 & 25258.0 & 15823.0 & 1.0 \\
betas[3] & 0.076 & 0.116 & -0.144 & 0.293 & 0.001 & 0.001 & 25471.0 & 15787.0 & 1.0 \\
betas[4] & -0.030 & 0.131 & -0.271 & 0.220 & 0.001 & 0.001 & 24966.0 & 15852.0 & 1.0 \\
betas[5] & -0.069 & 0.129 & -0.306 & 0.178 & 0.001 & 0.001 & 23914.0 & 15440.0 & 1.0 \\
betas[6] & 0.028 & 0.113 & -0.180 & 0.244 & 0.001 & 0.001 & 26766.0 & 15221.0 & 1.0 \\
sigma & 1.033 & 0.077 & 0.892 & 1.181 & 0.001 & 0.000 & 21031.0 & 13627.0 & 1.0 \\
\bottomrule
\end{tabular}
\label{tab:summary-effectiveness-stochastic}
\end{table}

\begin{table}[H]
\centering
\caption{Summary of posterior distributions for the parameters in the Bayesian model for \textbf{prediction accuracy} in the setting of playing against \textbf{dynamic bots}.}
\begin{tabular}{lcccccccccc}
\toprule
 & mean & sd & hdi\_3\% & hdi\_97\% & mcse\_mean & mcse\_sd & ess\_bulk & ess\_tail & r\_hat \\
\midrule
alpha & 0.000 & 0.105 & -0.200 & 0.195 & 0.001 & 0.001 & 27065.0 & 15234.0 & 1.0 \\
betas[0] & -0.098 & 0.117 & -0.308 & 0.131 & 0.001 & 0.001 & 23668.0 & 15874.0 & 1.0 \\
betas[1] & 0.040 & 0.131 & -0.198 & 0.297 & 0.001 & 0.001 & 23325.0 & 15570.0 & 1.0 \\
betas[2] & 0.070 & 0.112 & -0.145 & 0.279 & 0.001 & 0.001 & 26515.0 & 16378.0 & 1.0 \\
betas[3] & 0.003 & 0.116 & -0.217 & 0.221 & 0.001 & 0.001 & 28278.0 & 15693.0 & 1.0 \\
betas[4] & 0.166 & 0.129 & -0.075 & 0.409 & 0.001 & 0.001 & 26904.0 & 15537.0 & 1.0 \\
betas[5] & -0.032 & 0.129 & -0.277 & 0.202 & 0.001 & 0.001 & 25533.0 & 16145.0 & 1.0 \\
betas[6] & -0.065 & 0.113 & -0.280 & 0.143 & 0.001 & 0.001 & 27082.0 & 16197.0 & 1.0 \\
sigma & 1.036 & 0.078 & 0.898 & 1.189 & 0.001 & 0.000 & 22124.0 & 14574.0 & 1.0 \\
\bottomrule
\end{tabular}
\label{tab:summary-pred-accuracy-stochastic}
\end{table}

\begin{table}[H]
\centering
\caption{Summary of posterior distributions for the parameters in the Bayesian model for estimating the \textbf{decision effectiveness} in the RPS game against \textbf{human opponents}.}
\begin{tabular}{lcccccccccc}
\toprule
 & mean & sd & hdi\_3\% & hdi\_97\% & mcse\_mean & mcse\_sd & ess\_bulk & ess\_tail & r\_hat \\
\midrule
alpha & 0.000 & 0.064 & -0.121 & 0.120 & 0.000 & 0.000 & 31692.0 & 14913.0 & 1.0 \\
betas[0] & 0.087 & 0.070 & -0.041 & 0.222 & 0.000 & 0.000 & 28952.0 & 15613.0 & 1.0 \\
betas[1] & -0.033 & 0.084 & -0.191 & 0.127 & 0.001 & 0.001 & 26626.0 & 16072.0 & 1.0 \\
betas[2] & 0.023 & 0.066 & -0.101 & 0.150 & 0.000 & 0.000 & 29489.0 & 15506.0 & 1.0 \\
betas[3] & 0.043 & 0.072 & -0.085 & 0.186 & 0.000 & 0.000 & 30931.0 & 16411.0 & 1.0 \\
betas[4] & 0.056 & 0.093 & -0.119 & 0.229 & 0.001 & 0.001 & 25591.0 & 14694.0 & 1.0 \\
betas[5] & 0.018 & 0.077 & -0.122 & 0.164 & 0.000 & 0.001 & 28323.0 & 15843.0 & 1.0 \\
betas[6] & 0.004 & 0.066 & -0.119 & 0.126 & 0.000 & 0.000 & 29198.0 & 15003.0 & 1.0 \\
sigma & 1.010 & 0.046 & 0.927 & 1.099 & 0.000 & 0.000 & 26969.0 & 14395.0 & 1.0 \\
\bottomrule
\end{tabular}
\label{tab:summary-effectiveness-human}
\end{table}

\begin{table}[H]
\centering
\caption{Summary of posterior distributions for the parameters in the Bayesian model for \textbf{prediction accuracy} in the setting of playing against \textbf{human opponents}.}
\begin{tabular}{lcccccccccc}
\toprule
 & mean & sd & hdi\_3\% & hdi\_97\% & mcse\_mean & mcse\_sd & ess\_bulk & ess\_tail & r\_hat \\
\midrule
alpha & 0.000 & 0.063 & -0.114 & 0.122 & 0.000 & 0.000 & 30693.0 & 14999.0 & 1.0 \\
betas[0] & 0.034 & 0.070 & -0.102 & 0.159 & 0.000 & 0.000 & 28476.0 & 16523.0 & 1.0 \\
betas[1] & 0.005 & 0.084 & -0.151 & 0.166 & 0.001 & 0.001 & 27670.0 & 15420.0 & 1.0 \\
betas[2] & -0.021 & 0.066 & -0.148 & 0.098 & 0.000 & 0.000 & 28790.0 & 16016.0 & 1.0 \\
betas[3] & 0.044 & 0.072 & -0.091 & 0.178 & 0.000 & 0.000 & 27889.0 & 16422.0 & 1.0 \\
betas[4] & 0.093 & 0.092 & -0.079 & 0.265 & 0.001 & 0.000 & 24560.0 & 15056.0 & 1.0 \\
betas[5] & 0.018 & 0.077 & -0.128 & 0.163 & 0.000 & 0.001 & 31717.0 & 15921.0 & 1.0 \\
betas[6] & -0.065 & 0.066 & -0.189 & 0.061 & 0.000 & 0.000 & 28058.0 & 15923.0 & 1.0 \\
sigma & 1.008 & 0.046 & 0.920 & 1.092 & 0.000 & 0.000 & 26923.0 & 14717.0 & 1.0 \\
\bottomrule
\end{tabular}
\label{tab:summary-pred-accuracy-human}
\end{table}

\paragraph{Static bots.} Tables \ref{tab:summary-effectiveness-static} and \ref{tab:summary-pred-accuracy-static} present the fitting parameters for the staic bots setting.

\begin{figure}[htpb]
\vspace{-1ex}
\centering
\begin{subfigure}[b]{0.5\linewidth}
    \includegraphics[width=\linewidth]{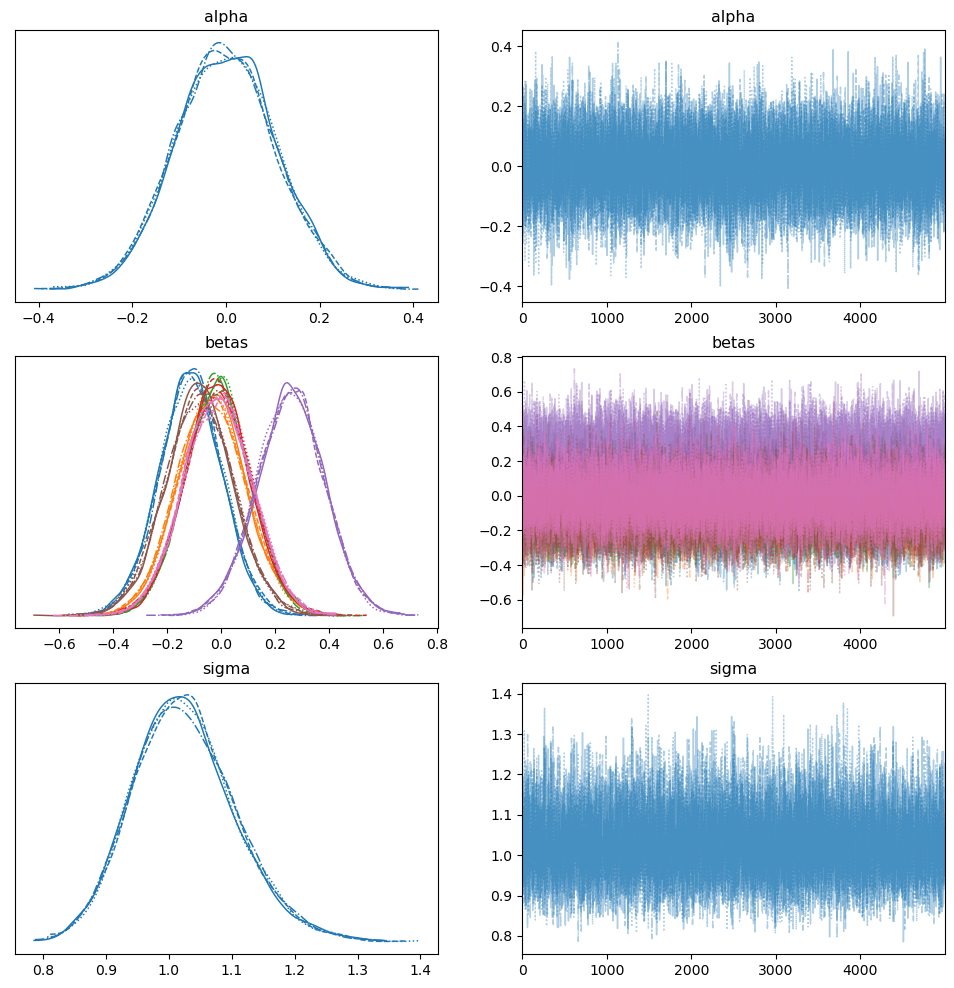}
    \caption{Decision Effectiveness.}
    \label{fig:trace-effectiveness-static}
\end{subfigure}%
\begin{subfigure}[b]{0.5\linewidth}
    \includegraphics[width=\linewidth]{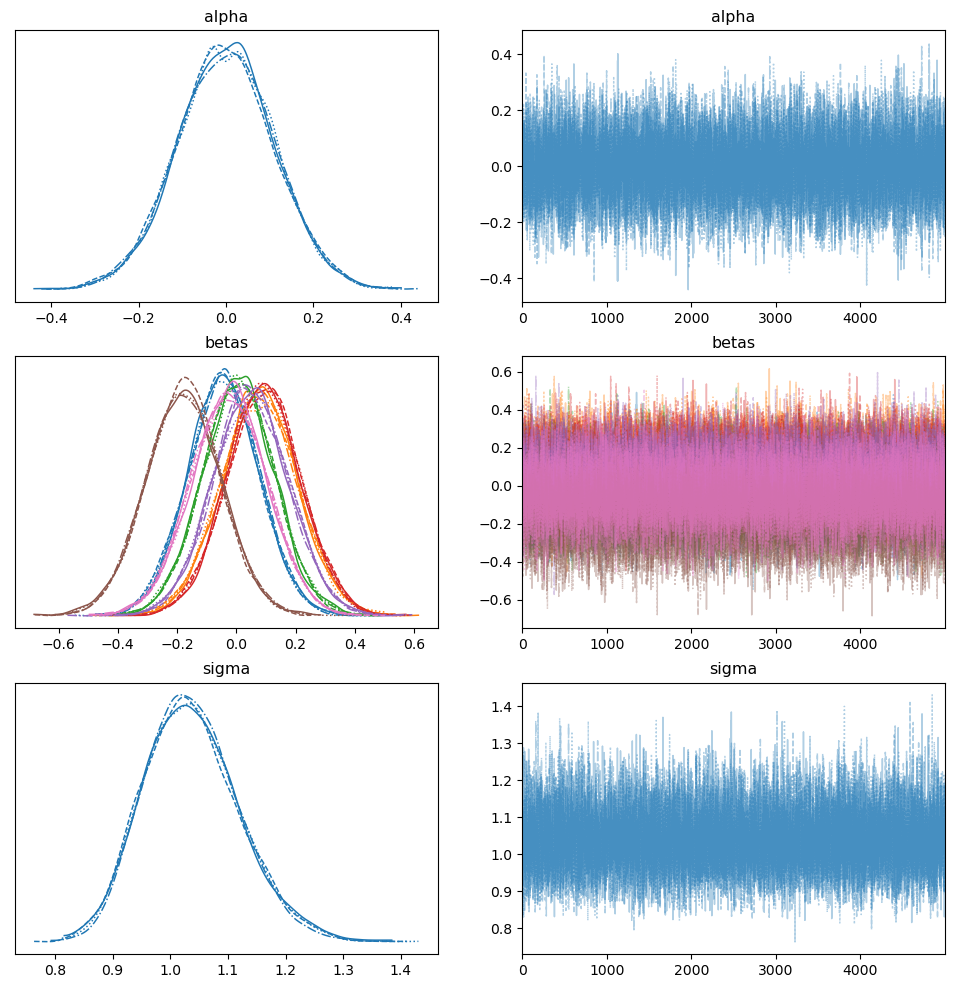}
    \caption{Prediction Accuracy.}
    \label{fig:trace-pred-acc-static}
\end{subfigure}
\vspace{-1ex}
\caption{Trace of the sampled values in the Bayesian model for static bot setting.}
\label{fig:combined-trace-plots-static}
\end{figure}


\paragraph{Dynamic bots.} Tables \ref{tab:summary-effectiveness-stochastic} and \ref{tab:summary-pred-accuracy-stochastic} present the fitting parameters for the dynamic bots setting.

\begin{figure}[htpb]
\vspace{-1ex}
\centering
\begin{subfigure}[b]{0.5\linewidth}
    \includegraphics[width=\linewidth]{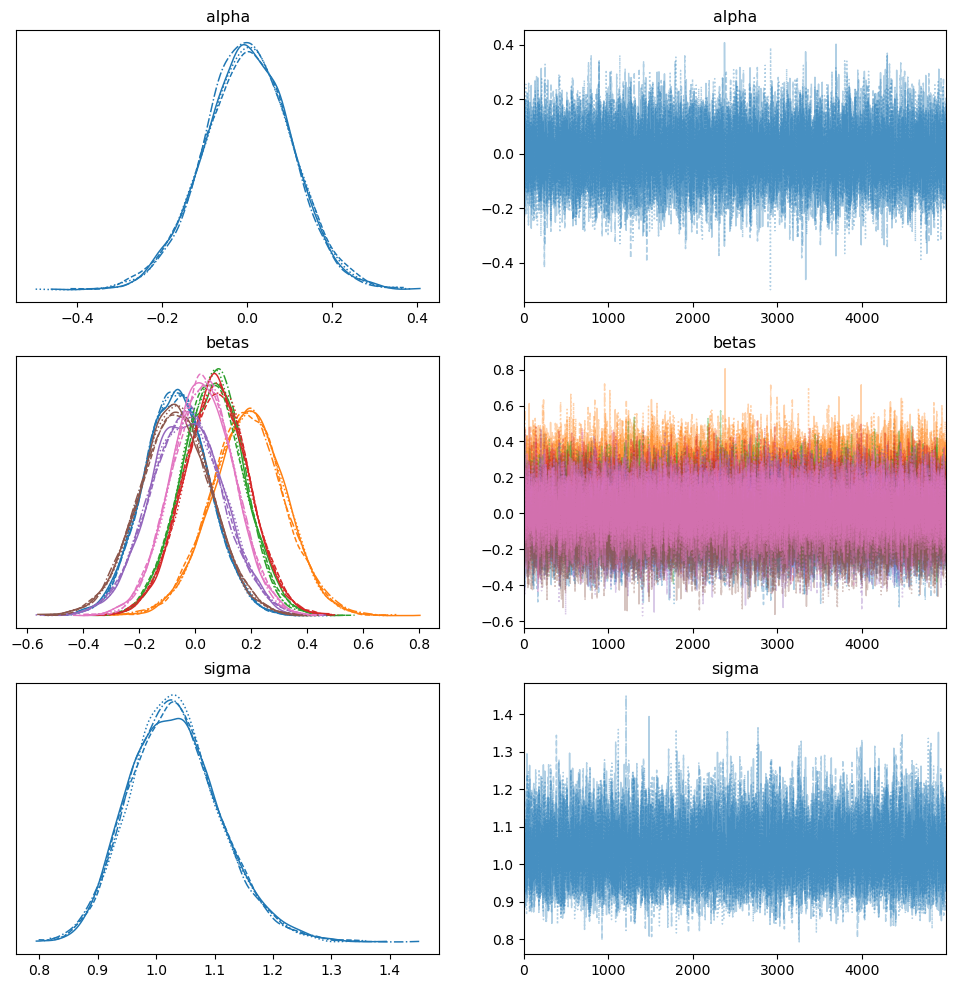}
    \caption{Decision Effectiveness.}
    \label{fig:trace-effectiveness-stochastic}
\end{subfigure}%
\begin{subfigure}[b]{0.5\linewidth}
    \includegraphics[width=\linewidth]{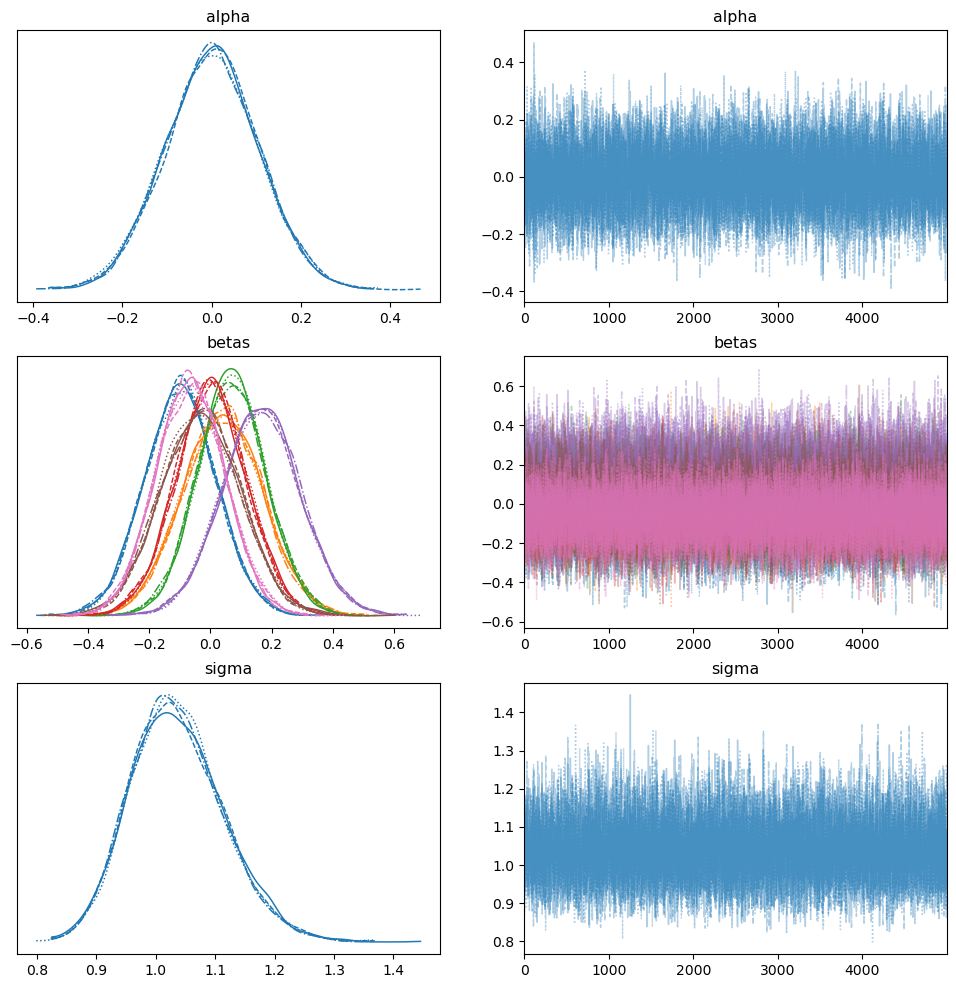}
    \caption{Prediction Accuracy.}
    \label{fig:trace-pred-acc-stochastic}
\end{subfigure}
\vspace{-1ex}
\caption{Trace of the sampled values in the Bayesian model for dynamic bot setting.}
\label{fig:combined-trace-plots-stochastic}
\end{figure}


\paragraph{Human Players.} Tables \ref{tab:summary-effectiveness-human} and \ref{tab:summary-pred-accuracy-human} present the fitting parameters for the setting of playing against other human players.

\begin{figure}[htpb]
\vspace{-1ex}
\centering
\begin{subfigure}[b]{0.5\linewidth}
    \includegraphics[width=\linewidth]{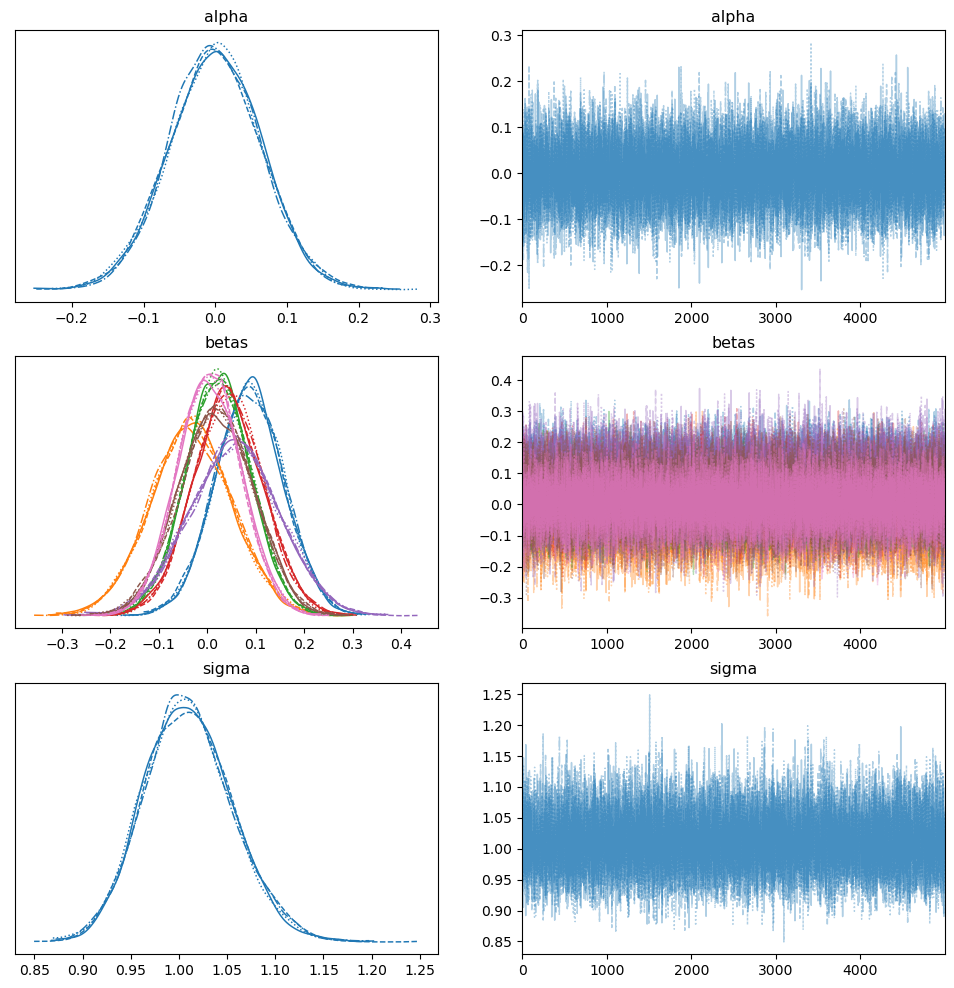}
    \caption{Decision Effectiveness.}
    \label{fig:trace-effectiveness-human}
\end{subfigure}%
\begin{subfigure}[b]{0.5\linewidth}
    \includegraphics[width=\linewidth]{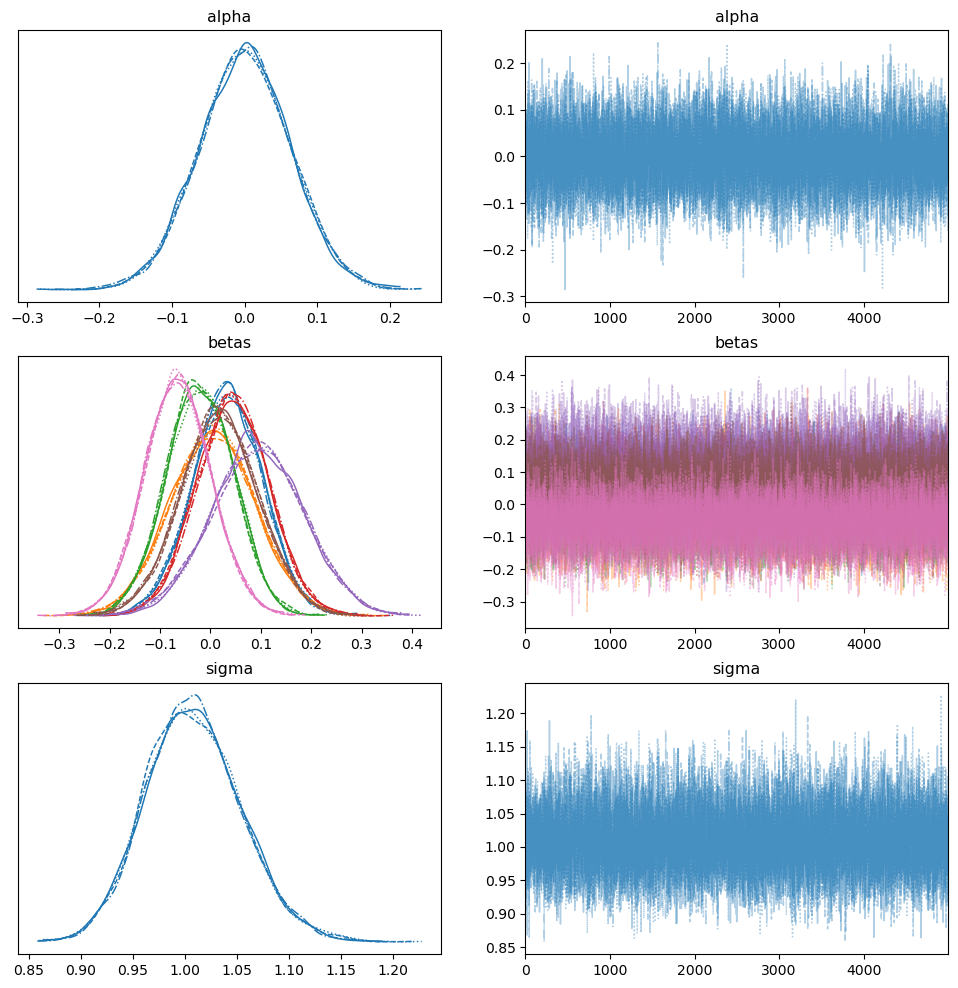}
    \caption{Prediction Accuracy.}
    \label{fig:trace-pred-acc-human}
\end{subfigure}
\vspace{-1ex}
\caption{Trace of the sampled values in the Bayesian model for human player setting.}
\label{fig:combined-trace-plots-human}
\end{figure}

\clearpage

\subsection{Results of Structural Equation Models Analysis} \label{appendix:sem}
We used the SEM package, \texttt{semopy}, to build the SEM models reported in Section~\ref{subsec:rq4}. The model variables in \texttt{semopy} syntax are as follows:
\begin{itemize}
    \item $x_0$: Interpersonal Reactivity Index
    \item $x_1$: Reading the Mind in the Eyes test scores (RME)
    \item $x_2$: Experiential Inventory
    \item $x_3$: Rational Inventory
    \item $x_4$: Recursive Thinking
    \item $x_5$: Spatial Reasoning
    \item $x_6$: Perspective Taking
    \item $y_1$: Decision Effectiveness
    \item $y_2$: Prediction Accuracy
\end{itemize}

\subsubsection{Static-strategy Bot Opponent}
\begin{lstlisting}[language=Python]
desc = '''
factor1 =~ x5 + x6
factor2 =~ x0 + x1
y1 ~ b1*factor1 + b2*factor2
'''
mod = Model(desc)
res_opt = mod.fit(df_data)
estimates = mod.inspect()
print(estimates)
\end{lstlisting}

\begin{lstlisting}
      lval  op     rval  Estimate  Std. Err   z-value   p-value
0        x5   ~  factor1  1.000000         -         -         -
1        x6   ~  factor1  1.120063  0.496415  2.256301  0.024052
2        x0   ~  factor2  1.000000         -         -         -
3        x1   ~  factor2  2.008080  3.080672  0.651832   0.51451
4        y1   ~  factor1 -0.006262  0.048128 -0.130121  0.896471
5        y1   ~  factor2  0.012933  0.046032  0.280956  0.778744
6   factor1  ~~  factor1  0.369032  0.197889  1.864843  0.062203
7   factor1  ~~  factor2  0.124660  0.199045  0.626292  0.531123
8   factor2  ~~  factor2  0.250308  0.439469  0.569569   0.56897
9        x0  ~~       x0  5.124539  0.858147  5.971633       0.0
10       x1  ~~       x1  0.000000  1.469015       0.0       1.0
11       x5  ~~       x5  0.627507  0.185566  3.381594  0.000721
12       x6  ~~       x6  0.546227  0.216351  2.524726  0.011579
13       y1  ~~       y1  0.032771  0.004971  6.592929       0.0
\end{lstlisting}

\begin{figure}[!htpb]
\vspace{-1ex}
\begin{center}
\includegraphics[width=.8\linewidth]{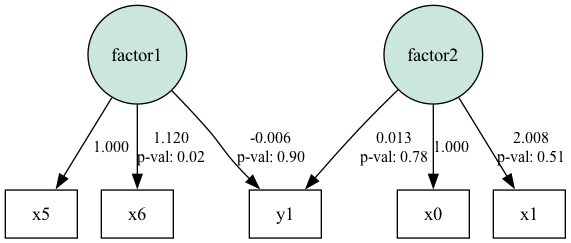} 
\caption{SEM for estimating the effectiveness in the static bot condition.
\label{fig:semopy-effectiveness-static}
} 
\end{center}
\vspace{-1ex}
\end{figure}

\begin{lstlisting}[language=Python]
desc = '''
factor1 =~ x5 + x6
factor2 =~ x0 + x1
y2 ~ b1*factor1 + b2*factor2
'''
mod2 = Model(desc)
res_opt2 = mod2.fit(df_data)
estimates2 = mod2.inspect()
print(estimates2)
\end{lstlisting}

\begin{lstlisting}
      lval  op     rval      Estimate  Std. Err   z-value   p-value
0        x5   ~  factor1  1.000000e+00         -         -         -
1        x6   ~  factor1  9.689223e-01   0.39573   2.44844  0.014348
2        x0   ~  factor2  1.000000e+00         -         -         -
3        x1   ~  factor2  2.010004e+00  2.983586  0.673687   0.50051
4        y2   ~  factor1 -6.900701e-02  0.065648 -1.051159  0.293185
5        y2   ~  factor2  6.376930e-02   0.06559   0.97224  0.330931
6   factor1  ~~  factor1  4.264159e-01   0.20969  2.033552  0.041997
7   factor1  ~~  factor2  1.334646e-01  0.205424  0.649705  0.515883
8   factor2  ~~  factor2  2.497708e-01  0.427887  0.583731  0.559401
9        x0  ~~       x0  5.131444e+00  0.853259  6.013931       0.0
10       x1  ~~       x1  7.396278e-17  1.415338       0.0       1.0
11       x5  ~~       x5  5.703585e-01  0.189964  3.002458  0.002678
12       x6  ~~       x6  6.084669e-01  0.183644  3.313298  0.000922
13       y2  ~~       y2  4.579621e-02  0.007306  6.267953       0.0
\end{lstlisting}

\begin{figure}[!htpb]
\vspace{-1ex}
\begin{center}
\includegraphics[width=.8\linewidth]{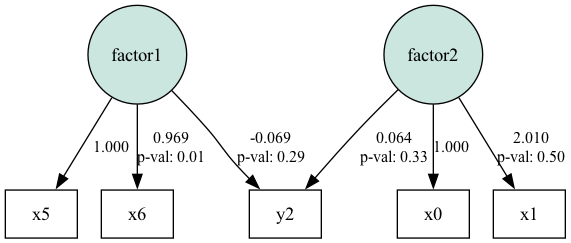} 
\caption{SEM for the prediction accuracy in the static bot condition.
\label{fig:semopy-pred-accuracy-static}
} 
\end{center}
\vspace{-1ex}
\end{figure}

\subsubsection{Dynamic-strategy Bot Opponent}
\begin{lstlisting}[language=Python]
desc = '''
factor1 =~ x1 + x4 + x5
factor2 =~ x0 
y1 ~ b1*factor1 + b2*factor2
'''
mod = Model(desc)
res_opt = mod.fit(df_data)
estimates = mod.inspect()
print(estimates)
\end{lstlisting}

\begin{lstlisting}
WARNING:root:Fisher Information Matrix is not PD.Moore-Penrose inverse will be used instead of Cholesky decomposition. See 10.1109/TSP.2012.2208105.
WARNING:root:Fisher Information Matrix is not PD.Moore-Penrose inverse will be used instead of Cholesky decomposition. See 10.1109/TSP.2012.2208105.
       lval  op     rval  Estimate  Std. Err   z-value   p-value
0        x1   ~  factor1  1.000000         -         -         -
1        x4   ~  factor1  0.990470  0.226696  4.369165  0.000012
2        x5   ~  factor1  0.904417  0.209397  4.319153  0.000016
3        x0   ~  factor2  1.000000         -         -         -
4        y1   ~  factor1  0.353741  0.080142   4.41392   0.00001
5        y1   ~  factor2 -0.301090  0.056244 -5.353304       0.0
6   factor1  ~~  factor1  0.477687  0.156782  3.046813  0.002313
7   factor1  ~~  factor2  0.534265  0.217285  2.458818   0.01394
8   factor2  ~~  factor2  0.617073  0.447856  1.377837  0.168254
9        x0  ~~       x0  5.061276  0.772358  6.553017       0.0
10       x1  ~~       x1  0.524592  0.122959  4.266399   0.00002
11       x4  ~~       x4  0.532883  0.122392  4.353897  0.000013
12       x5  ~~       x5  0.582472  0.116617  4.994734  0.000001
13       y1  ~~       y1  0.011900  0.008537  1.394002  0.163317
\end{lstlisting}

\begin{figure}[!htpb]
\vspace{-1ex}
\begin{center}
\includegraphics[width=.8\linewidth]{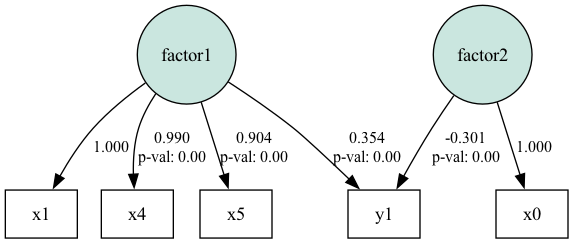} 
\caption{SEM for estimating the effectiveness in the dynamic bot condition.
\label{fig:semopy-effectiveness-stochastic}
} 
\end{center}
\vspace{-1ex}
\end{figure}

\begin{lstlisting}[language=Python]
desc = '''
factor1 =~ x1 + x4 + x5
factor2 =~ x0 
y2 ~ b1*factor1 + b2*factor2
'''
mod2 = Model(desc)
res_opt2 = mod2.fit(df_data)
estimates2 = mod2.inspect()
print(estimates2)
\end{lstlisting}

\begin{lstlisting}
WARNING:root:Fisher Information Matrix is not PD.Moore-Penrose inverse will be used instead of Cholesky decomposition. See 10.1109/TSP.2012.2208105.
WARNING:root:Fisher Information Matrix is not PD.Moore-Penrose inverse will be used instead of Cholesky decomposition. See 10.1109/TSP.2012.2208105.
       lval  op     rval  Estimate  Std. Err   z-value   p-value
0        x1   ~  factor1  1.000000         -         -         -
1        x4   ~  factor1  1.039938  0.237237  4.383539  0.000012
2        x5   ~  factor1  0.931200  0.214772  4.335762  0.000015
3        x0   ~  factor2  1.000000         -         -         -
4        y2   ~  factor1  0.377989  0.150747  2.507441  0.012161
5        y2   ~  factor2 -0.318865  0.080779 -3.947366  0.000079
6   factor1  ~~  factor1  0.454770  0.152258  2.986846  0.002819
7   factor1  ~~  factor2  0.511924  0.212321  2.411078  0.015905
8   factor2  ~~  factor2  0.650937  0.438083  1.485876  0.137312
9        x0  ~~       x0  5.027187  0.749825  6.704484       0.0
10       x1  ~~       x1  0.547662  0.121428  4.510164  0.000006
11       x4  ~~       x4  0.509792  0.123299  4.134581  0.000036
12       x5  ~~       x5  0.579041  0.116418  4.973827  0.000001
13       y2  ~~       y2  0.006963  0.009524  0.731118  0.464707
\end{lstlisting}

\begin{figure}[!htpb]
\vspace{-1ex}
\begin{center}
\includegraphics[width=.8\linewidth]{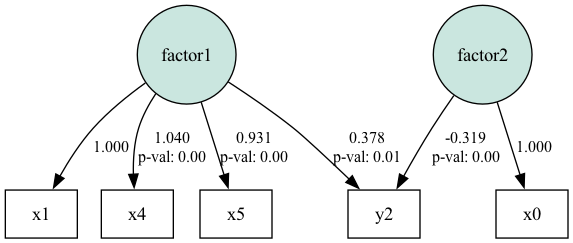} 
\caption{SEM for the prediction accuracy in the dynamic bot condition.
\label{fig:semopy-pred-accuracy-stochastic}
} 
\end{center}
\vspace{-1ex}
\end{figure}

\subsubsection{Human Player Opponent}
\begin{lstlisting}[language=Python]
desc = '''
factor1 =~ x1 + x4 + x5
factor2 =~ x3
y1 ~ b1*factor1 + b2*factor2
'''
mod = Model(desc)
res_opt = mod.fit(df_data)
estimates = mod.inspect()
print(estimates)
\end{lstlisting}

\begin{lstlisting}
WARNING:root:Fisher Information Matrix is not PD.Moore-Penrose inverse will be used instead of Cholesky decomposition. See 10.1109/TSP.2012.2208105.
WARNING:root:Fisher Information Matrix is not PD.Moore-Penrose inverse will be used instead of Cholesky decomposition. See 10.1109/TSP.2012.2208105.
       lval  op     rval  Estimate  Std. Err    z-value   p-value
0        x1   ~  factor1  1.000000         -          -         -
1        x4   ~  factor1  1.168249  0.126591   9.228513       0.0
2        x5   ~  factor1  0.791995  0.097411   8.130441       0.0
3        x3   ~  factor2  1.000000         -          -         -
4        y1   ~  factor1  0.175640  0.023114   7.598816       0.0
5        y1   ~  factor2 -0.299923  0.016466  -18.21495       0.0
6   factor1  ~~  factor1  0.535602  0.092612   5.783284       0.0
7   factor1  ~~  factor2  0.291792   0.05883   4.959883  0.000001
8   factor2  ~~  factor2  0.140330  0.057362   2.446378   0.01443
9        x1  ~~       x1  0.458013  0.063524   7.210115       0.0
10       x3  ~~       x3  0.868167  0.083616  10.382753       0.0
11       x4  ~~       x4  0.266960  0.069383   3.847624  0.000119
12       x5  ~~       x5  0.665082  0.068052   9.773078       0.0
13       y1  ~~       y1  0.012965  0.002251   5.760646       0.0
\end{lstlisting}

\begin{figure}[!htpb]
\vspace{-1ex}
\begin{center}
\includegraphics[width=.8\linewidth]{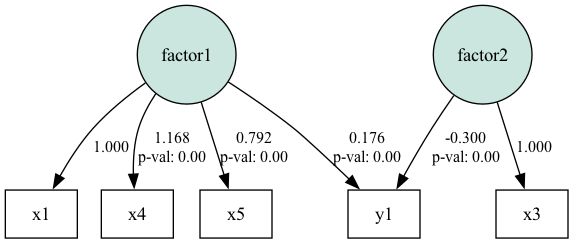} 
\caption{SEM for estimating the effectiveness in the human opponent condition.
\label{fig:semopy-effectiveness-human}
} 
\end{center}
\vspace{-1ex}
\end{figure}

\begin{lstlisting}[language=Python]
desc = '''
factor1 =~ x1 + x4 + x5
factor2 =~ x3
y2 ~ b1*factor1 + b2*factor2
'''
mod2 = Model(desc)
res_opt2 = mod2.fit(df_data)
estimates2 = mod2.inspect()
print(estimates2)
\end{lstlisting}

\begin{lstlisting}
WARNING:root:Fisher Information Matrix is not PD.Moore-Penrose inverse will be used instead of Cholesky decomposition. See 10.1109/TSP.2012.2208105.
WARNING:root:Fisher Information Matrix is not PD.Moore-Penrose inverse will be used instead of Cholesky decomposition. See 10.1109/TSP.2012.2208105.
       lval  op     rval  Estimate  Std. Err    z-value   p-value
0        x1   ~  factor1  1.000000         -          -         -
1        x4   ~  factor1  1.169669  0.126224   9.266635       0.0
2        x5   ~  factor1  0.792144  0.097355   8.136638       0.0
3        x3   ~  factor2  1.000000         -          -         -
4        y2   ~  factor1  0.005253  0.030749   0.170845  0.864346
5        y2   ~  factor2  0.033094  0.049179   0.672939  0.500986
6   factor1  ~~  factor1  0.535335  0.092457   5.790099       0.0
7   factor1  ~~  factor2  0.291462   0.05877   4.959379  0.000001
8   factor2  ~~  factor2  0.313275  0.045471   6.889499       0.0
9        x1  ~~       x1  0.458633  0.063325   7.242556       0.0
10       x3  ~~       x3  0.694892  0.045763  15.184742       0.0
11       x4  ~~       x4  0.265901  0.069063   3.850112  0.000118
12       x5  ~~       x5  0.665145  0.068007   9.780571       0.0
13       y2  ~~       y2  0.016803  0.001569  10.711336       0.0
\end{lstlisting}

\begin{figure}[!htpb]
\vspace{-1ex}
\begin{center}
\includegraphics[width=.8\linewidth]{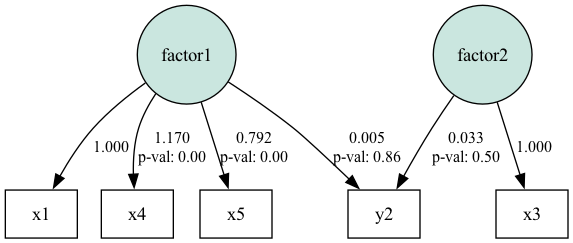} 
\caption{SEM for the prediction accuracy in the human opponent condition.
\label{fig:semopy-pred-accuracy-human}
} 
\end{center}
\vspace{-1ex}
\end{figure}

\end{document}